\documentclass[aps,prb,twocolumn,floatfix,footinbib,showpacs]{revtex4-1}
\usepackage{graphicx}
\usepackage{amsfonts,amsmath,amssymb}
\usepackage{amsthm}
\usepackage{dsfont,bm}
\usepackage{color}
\usepackage{soul} 
\usepackage{amsbsy}
\usepackage[colorlinks=true,linkcolor=blue,pagecolor=blue,filecolor=blue,menucolor=blue,urlcolor=blue,citecolor=blue,anchorcolor=blue]{hyperref}%

\usepackage{mathbbol}

\usepackage{sidecap}

\newcommand{\bsigma}{{\boldsymbol \sigma}}

\begin{document}

\title{Critical Supercurrent and $\varphi_0$-State for Probing a Persistent Spin Helix}

\author{Mohammad Alidoust}
\affiliation{Department of Physics, K.N. Toosi University of Technology, Tehran 15875-4416, Iran}

\date{\today}

\begin{abstract}
We theoretically study the profile of a supercurrent in two-dimensional Josephson junctions with Rashba-Dresselhaus spin-orbit interaction (RDSOI) in the presence of a Zeeman field. Two types of RDSOIs are considered that might be accessible in $\rm GaAs$ quantum wells and zinc-blende materials. Through investigating self-biased supercurrent (so called $\varphi_0$-Josephson state), we obtain explicit expressions for the functionality of the $\varphi_0$ state with respect to RDSOI parameters ($\alpha,\beta$) and in-plane Zeeman field components ($h_x,h_y$). Our findings reveal that, when the chemical potential ($\mu$) is high enough compared to the energy gap ($\Delta$) in superconducting electrodes, i.e., $\mu \gg \Delta$, RSOI and DSOI with equal strengths ($|\alpha|=|\beta|$) cause vanishing $\varphi_0$ state independent of magnetization and the type of RDSOI. A Zeeman field with unequal components, i.e., $|h_x|\neq |h_y|$, however, can counteract and nullify the destructive impact of equal-strength RDSOIs (for one type only), where $\mu\sim\Delta$, although $|h_x|= |h_y|$ can still eliminate the $\varphi_0$ state. Remarkably, in the $\mu\sim\Delta$ limit, the $\varphi_0$ state is proportional to the multiplication of both components of an in-plane Zeeman field, i.e., $h_xh_y$, which is absent in the $\mu \gg \Delta$ limit. Furthermore, our results of critical supercurrents demonstrate that the persistent spin helices can be revealed in a high enough chemical potential regime $\mu\gg \Delta$, while an opposite regime, i.e., $\mu\sim\Delta$, introduces an adverse effect. In the ballistic regime, the ``maximum'' of the critical supercurrent occurs at $|\alpha|=|\beta|$ and the Zeeman field can boost this feature. The presence of disorder and nonmagnetic impurities change this picture drastically so the ``minimum'' of the critical supercurrent occurs at and around the symmetry lines $|\alpha|=|\beta|$. We show that the signature of persistent spin helices explored in disordered systems originate from the competition of short-range spin-singlet and long-range spin-triplet supercurrent components. 
Our study uncovers delicate details of how the interplay of RDSOI and a Zeeman field manifests in the $\varphi_0$ state and critical supercurrent. Relying on the fact that the $\varphi_0$ state is accessible regardless of the amount of nonmagnetic impurities and disorder, our results can provide guidelines for future experiments to confirm the presence of persistent spin helices, determine the type of SOI, and reliably extract SOI parameters in a system, which might be helpful in devising spin-orbit-coupled spintronics devices and ultra sensitive spin-transistor technologies. 
\end{abstract}
\maketitle

\section{Introduction}

The quantum response of a system to motives can be highly influenced by the electron spin and orbital degrees of freedom as well as their interaction. This fact has triggered immense interest to, first, shed light on various aspects of spin-orbital interactions (SOIs) and, second, explore and propose practically feasible routes in utilizing SOI for spintronics devices \cite{I.Zutic,S.A.Wolf,I.Zutic2,I.Zutic3}. The spin-orbital-mediated interaction in a material can be either extrinsic or intrinsic, which removes spin degeneracy in the absence of magnetic field. The spin-dependent impurities can cause extrinsic SOI, providing only limited tuneable experimental knobs for controlling SOI. The intrinsic SOI can, however, originate from bulk inversion asymmetry (known as Dresselhaus spin splitting \cite{Dresselhaus}) or structure inversion symmetry due to confining potentials (known as [Bychkov-] Rashba spin splitting \cite{Rashba}). The intrinsic SOI can facilitate an externally controlled SOI by efficiently responding to the application of mechanical strain, electric field, or gate voltage, for instance \cite{Y.Ohno,Koralek,Walser}.   

\begin{SCfigure*}
\centering
\includegraphics[width=10.0cm,height=3.0cm, trim=1.3cm 6cm 1.0cm 6cm]{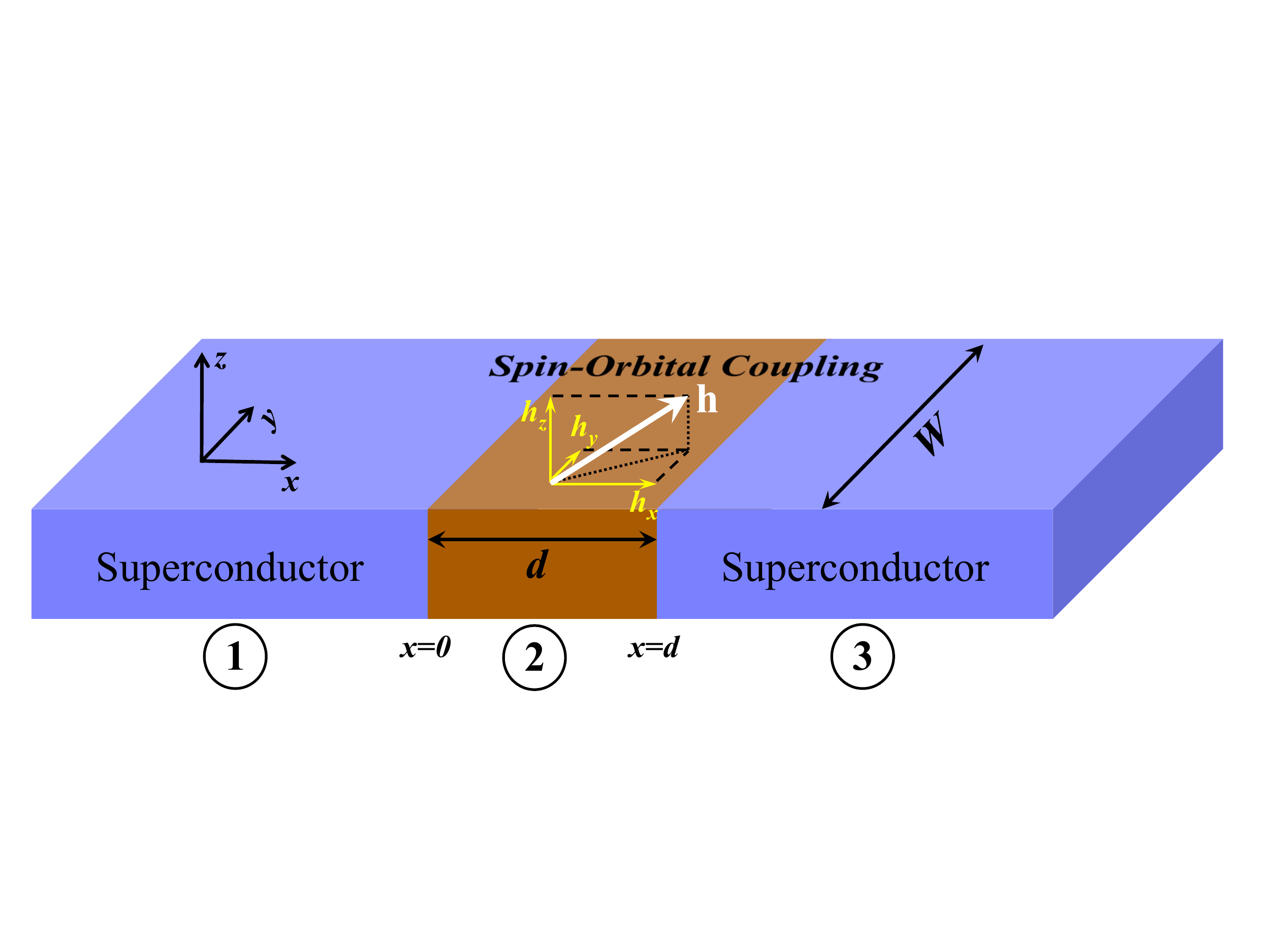}
\caption{\label{fig1} (Color online). The two-dimensional Josephson junction set up with a finite size width $W$ and thickness $d$. The junction is placed in the $xy$ plane so the interfaces are located at $x=0, d$ along the $y$ direction. The system can host different types of spin-orbital coupling in the presence of superconductivity and magnetization with an arbitrary orientation: $\textbf{h}=(h_x,h_y,h_z)$. To facilitate our discussion, we have labeled the three regions by $1,2,3$. }
\end{SCfigure*} 

The simultaneous existence of Rashba and Dresselhaus spin-orbit coupling (RSOC and DSOC) can result in fundamentally important phenomena such as persistent spin helix \cite{Schliemann,Ganichev,Y.Ohno,Koralek,Walser,Kammermeier,carlos_prl2016,carlos_prx2017,carlos_prl2019,Y.Feng}. This phenomenon occurs in certain directions of momentum space where spins oriented along these directions become insensitive to orbital field, and therefore the spin-splitting vanishes. In particular, the persistent spin helix allows for extremely long spin relaxation time and the generation of highly long-ranged spin-polarized currents (propagating over a long distance of the order of 8-25~$\mu$m) \cite{carlos_prl2016,carlos_prx2017,carlos_prl2019,alidoust1,alidoust2,H.Chakraborti}. A practical route traveled in recent years to determine the parameters of SOI in a system is transport measurements. This method has stimulated several theoretical and experimental works to provide a more realistic overview of Rashba-Dresselhaus spin-orbit interaction (RDSOI). The experiments eventually achieved a stretchable persistent spin helix, namely, a gate fine-tuned and continuously locked RDSOI at $|\alpha|=|\beta|$. The unidirectionality of RDSOI can be limited by the presence of cubic a DSOI term. Nevertheless, it has experimentally been found that the cubic Dresselhaus field \cite{J.J.Krich} should be highly small in $\rm GaAs$ quantum wells so $|\alpha|=|\beta|$ allows for 8-25$\mu$m long-distance communication. Hence, the persistent spin helix can reduce detergent spin dephasings and pave the way for designing diffusive spin field effect transistors. The well-known platforms for RDSOI-related phenomena are $\rm GaAs$ quantum wells and zinc-blende materials \cite{Schliemann,Ganichev}. In spite of continuous theoretical and experimental endeavors so far, the reliable experimental extraction of spin-orbit coupling (SOC) parameters is still elusive. Generally, providing an ideal situation in an experiment as theory assumes is highly challenging. There are several factors, such as detrimental impurities and unwanted defects, that influence adversely experimental results. Therefore, any experimentally observable SOI-related quantity, which is accessible regardless of the amount of nonmagnetic impurities and disorder (i.e., emerges in both ballistic and diffusive systems) is highly desired for conclusively determining the parameters of SOI.

When a ferromagnet is placed next to a $s$-wave superconductor, long-range spin superconducting correlations may arise due to the interplay of superconductivity and spatially textured spin at a close vicinity to the ferromagnet-superconductor interface \cite{fsf2,Buzdin2005,first,Halterman2007,Keizer2006,A.P.Mackenzie,I.Zutic2}. This proximity-induced phenomenon has stimulated numerous research works, ranging from critical temperature \cite{Keizer2006,fsf7,fsf42,halter_trip2,sff1,fsf5,halt2,half,bernard1,L.R.Tagirov} and density of states \cite{zep,zep2,zep3} to transport \cite{sfs4,Zdravkov,Antropov,Iovan,valve1,sff2,sfsf,A.A.Kamashev,M.G.Flokstra,A.A.Kamashev2,T.Vezin,jap_al} studies for characterizing and affirming their existence. Also, it has been theoretically shown that the interaction of SOI and $s$-wave superconducting can generate long-range spin triplet superconducting correlations \cite{Z.Niu,bergeret1,alidoust1,alidoust2,H.Chakraborti,N.Satchell,I.Zutic3,
 R.Beiranvand1,R.Beiranvand2,I.Martinez} and the accumulation of spin supercurrent at the edges of a finite-sized sample \cite{alidoust1,alidoust2}. Nevertheless, an experimentally clear-cut confirmation of these proximity spin superconducting correlations is still not achieved.      

The energy ground state of a conventional ferromagnetic Josephson junction with a uniform magnetization can always be found at two specific superconducting phase differences, i.e., $\varphi$=$0$ and $\pi$ \cite{A.G.Golubov,Ryazanov1,Ryazanov2,Fominov1,Takahashi,Eremin,
andrey2,andrey1,Rezaei,Hikino,Setiawan}. However, it is well understood that the interplay of SOC and magnetization can invalidate this picture. In this case, depending on parameter values and magnetization direction, the junction ground state can occur at any value of the phase difference, $\varphi=\varphi_0$, other than the $0$ and $\pi$. This class of Josephson effect is the so-called $\varphi_0$-Josephson state (a traditional Josephson effect with an extra $\varphi_0$ phase shift) \cite{A.Yu.Zyuzin,A.A.Reynoso,A.I.Buzdin,D.B.Szombati,zu1,zu2,zu3,I.V.Bobkova,
AlidoustWS2,AlidoustBP1,AlidoustBP2,herve,AlidoustWS1}. Due to the fundamentally important role that a $\varphi_0$ state can play in memory devices, recent theoretical and experimental efforts caused striking progress in observing $\varphi_0$ states using the surface of three-dimensional topological insulators, hosting strong SOI \citep{zu1,zu2,zu3,herve}. It has also theoretically been found that the $\varphi_0$ state is accessible in both ballistic and diffusive regimes independent of the presence of nonmagnetic impurities and disorder \cite{zu1,AlidoustWS2}. Nevertheless, a comprehensive study of how the $\varphi_0$ state depends on the components of a Zeeman field and RSOC-DSOC is still lacking in the literature.

Here, we study the $\varphi_0$ state driven by the interplay of a Zeeman field with generic in-plane orientation ($h_x, h_y, h_z$) and two types of Rashba-Dresselhaus SOCs (RDSOCs) with differing strengths [RSOC($\alpha$) $\neq$ DSOC($\beta$)]. To this end, a two-dimensional Josephson junction is considered and the profile of supercurrent flow across the junction in both the ballistic and diffusive regimes is determined when the junction is oriented along the $x$ axis (depicted in Fig. \ref{fig1}) and rotated by $90^\circ$ around the $z$ axis. The explicit expressions obtained for the $\varphi_0$ state illustrate that $|\alpha|$=$|\beta|$ eliminates the self-biased current when chemical potential is high enough compared to superconducting gap $\Delta$ inside the superconductor electrodes, i.e., $\mu\gg \Delta$. Decreasing $\mu$ to a low enough level, e.g., $\mu$ $\sim$ $\Delta$, the magnetization can retrieve the $\varphi_0$ state for one type of RDSOC although $\varphi_0$-state still vanishes in a certain regime of the Zeeman field, i.e., $|h_x|$=$|h_y|$. Interestingly, the $\varphi_0$ state directly depends on $h_xh_y$ terms, a feature which is absent in the $\mu\gg\Delta$ limit. Also, our study reveals that a low chemical potential $\mu$ $\sim$ $\Delta$ adversely impacts the signature of persistent spin helices on critical supercurrents. In a ballistic and short junction limit, where the spin-singlet supercurrent dominates, a large enough chemical potential $\mu\gg\Delta$ results in a tangible indication of persistent spin helices so the maximum of critical supercurrent occurs at and around $|\alpha|=|\beta|$ symmetry lines. Decomposing a supercurrent into its constituting components (spin-singlet, spin-triplets, and crossed terms), we demonstrate that in a long enough and diffusive junction, the spin-triplet components significantly enhance the supercurrent away from $|\alpha|=|\beta|$ symmetry lines. Therefore, in the diffusive regime, the presence of persistent spin helices the suppresses supercurrent at and around $|\alpha|=|\beta|$, providing an experimentally prominent and detectable evidence for both the stretchable persistent spin helix and spin-triplet supercurrent. 

Considering the accessibility of the $\varphi_0$ state in both the ballistic and diffusive systems, the expressions obtained for the $\varphi_0$ phase shift can provide a unique opportunity for experimentally extracting reliable parameter values for Rashba and Dresselhaus SOI parameters. Our theoretical findings can serve as guidance for examining the persistent spin helices discussed earlier. The distinctive influence of the spin-singlet and spin-triplet correlation on critical supercurrents found in this paper can serve as evidence for affirming the spin-triplet supercurrent.

The paper is organized as follows. In Sec. \ref{sec.theory}, we summarize theoretical frame works employed for studying a two-dimensional Josephson junction. The low-energy Hamiltonian, the Bogoliubov de Gennes approach, and charge current are described. Section \ref{sec.results} presents the main findings of the paper, divided into two subsections for the distinction between the self-biased supercurrent and critical supercurrent for various sets of parameters, considering the ballistic and diffusive systems. We finally give concluding remarks in Sec. \ref{sec.conclusion}.  

\begin{figure*}
\includegraphics[width=18.0cm,height=9.0cm]{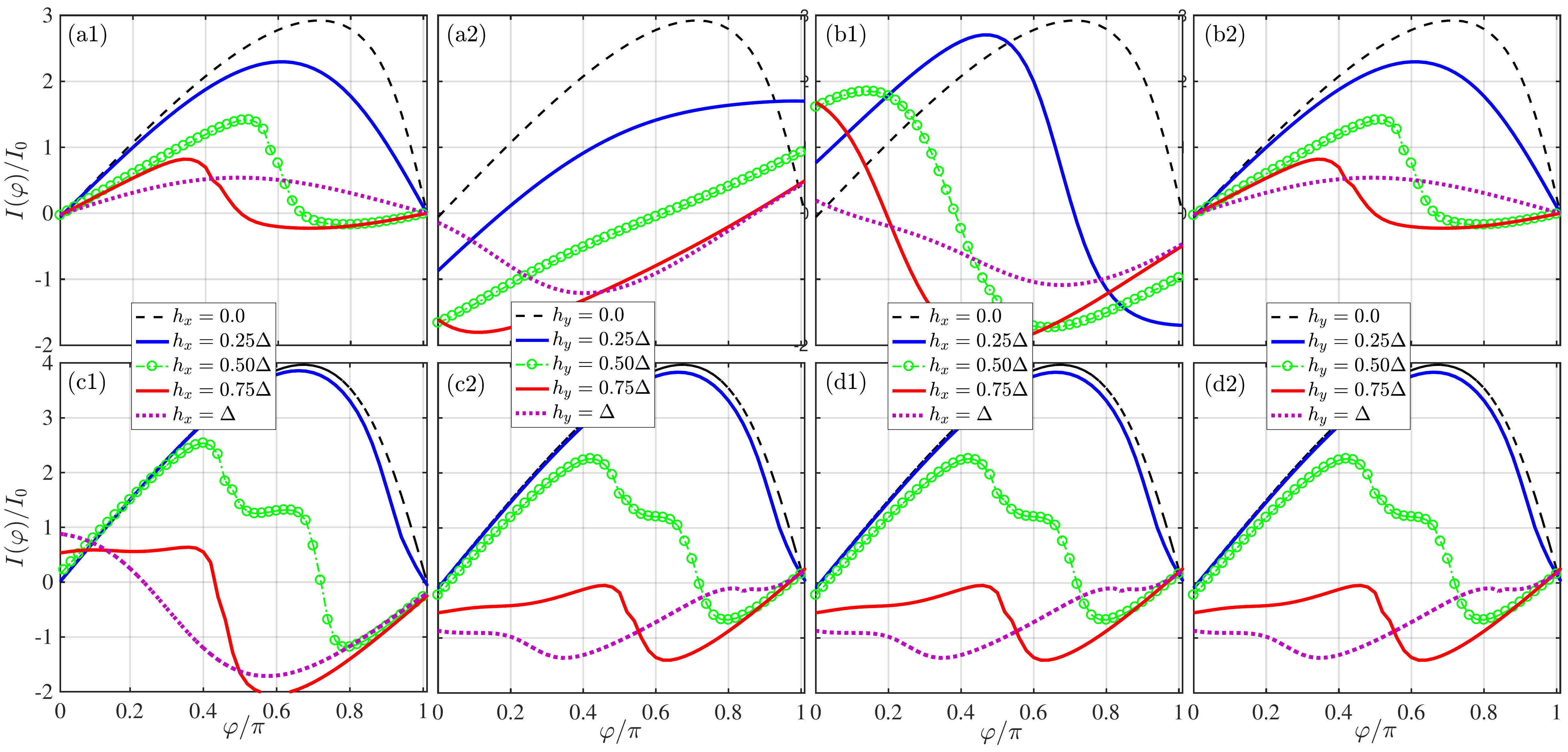}
\caption{\label{fig2} (Color online). Supercurrent as a function of superconducting phase difference, $I(\varphi)$, when the Josephson junction is oriented along the $x$ direction and spin-orbit interaction is $\boldsymbol{\eta}_\text{so}^\text{a}$. The supercurrent is plotted for various values of magnetization direction/intensity and the strength of spin-orbit interaction. In (a1)-(a2) we set $\alpha=1,\beta=0$, whereas in (b1)-(b2) $\alpha=0,\beta=1$. The spin-orbit parameters are set $\alpha=1,\beta=1$ in (c1)-(c2), while in (d1)-(d2) $\alpha=1,\beta=-1$. In the first and third columns, magnetization is oriented along the $x$ direction, i.e., $h_y=h_z=0, h_x\neq 0$ and in the second and fourth columns the orientation is aligned with the $y$ direction, i.e., $h_x=h_z=0, h_y\neq 0$.
}
\end{figure*}

\section{Theoretical approach}\label{sec.theory}

\subsection{Low-energy effective Hamiltonian}\label{sec.theory.hamilt}

The low-energy electronic properties of a noncentrosymmetric solid-state crystal, hosting spin-orbital-mediated interaction, can be described by an effective single-particle Hamiltonian
\begin{equation}
H =\frac{1}{2}\int d\mathbf{p}~ \hat{\psi}^{\dag}(\mathbf{p}) H(\mathbf{p})\hat{\psi}(\mathbf{p}),
\end{equation}
where 
\begin{eqnarray}\label{Hamil_so}
H(\mathbf{p})  =  \left[\frac{\mathbf{p}^2}{2m} +\bsigma\cdot \left(\boldsymbol{\eta}_\text{so} + \mathbf{h}\right)\right], 
\end{eqnarray}
in which $\mathbf{p}=(p_x,p_y,0)$ is momentum, $m$ is the effective mass of a charged particle, $\boldsymbol{\eta}_\text{so} $ is the SOI, $\mathbf{h}=(h_x,h_y,h_z)$ is the Zeeman energy corresponding to the magnetic field such that $|\mathbf{h}|= g\mu_B B$, $g$ is the g-factor of a charged carrier, $\mu_B$ is the Bohr magneton, and $B$ is the magnitude of the magnetic field. The field operator in spin-space can be expressed by $\hat{\psi}=(\psi_{\uparrow}, \psi_{\downarrow})^{\mathrm{T}}$ and $\bsigma=(\sigma_{x},\sigma_{y}, \sigma_{z})$ is a vector comprised of Pauli matrices. We use $\hbar = k_B= 1$ units throughout the paper.  

To simplify calculations, derive analytical expressions, and facilitate analysis, we consider two types of linearized SOI. 
\begin{subequations}\label{so}
\begin{align}\label{so1}
\boldsymbol{\eta} _\text{so}^\text{a} = \Big(\alpha p_y + \beta p_x, -\alpha p_x - \beta p_y, 0\Big),
\end{align}
\begin{align}\label{so2}
\boldsymbol{\eta}_\text{so}^\text{b} = \Big([\alpha + \beta] p_y, [\beta - \alpha] p_x, 0\Big),
\end{align}
\end{subequations}
in which $\alpha$ and $\beta$ are the Bychkov-Rashba and Dresselhaus velocities, respectively. As mentioned in the Introduction, the coefficients $\alpha$ and $\beta$ can be controlled by the application of electric field and mechanical strain. The two different types of RSOC and DSOC considered in this paper might be found in $\rm GaAs$ quantum wells and zinc-blende materials. The difference between $\boldsymbol{\eta} _\text{so}^\text{a} $ and $\boldsymbol{\eta} _\text{so}^\text{b} $ can originate from different crystallographic growth orientations. 
 
\begin{SCfigure*}
\centering
\includegraphics[width=14.0cm,height=9.0cm]{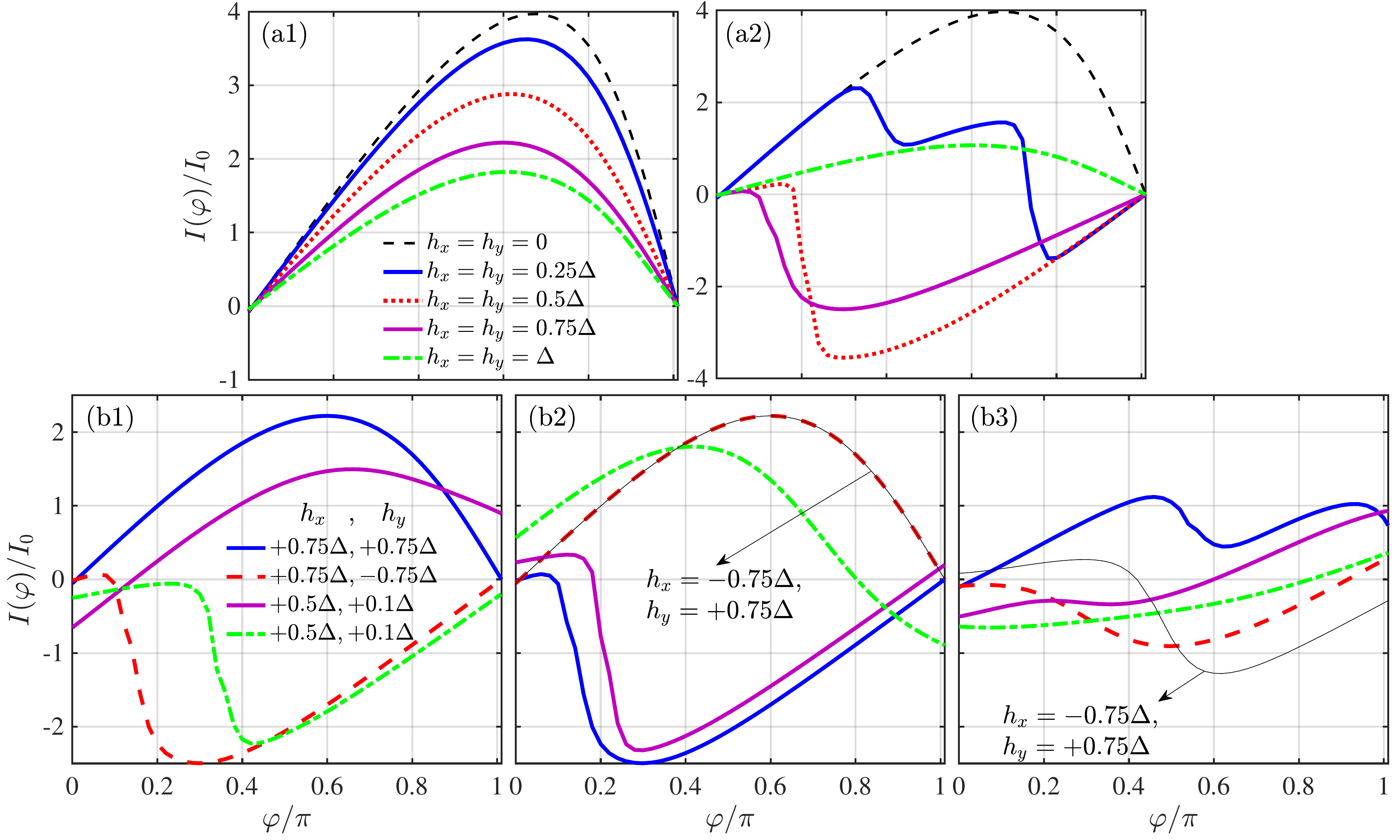}
\caption{\label{fig3} (Color online). Supercurrent as a function of superconducting phase difference when magnetization possesses two components in the plane of Josephson junction, i.e., $h_x\neq 0$ and $h_y\neq 0$. The junction is oriented along the $x$ direction and the spin-orbit interaction is described by $\boldsymbol{\eta}_\text{so}^\text{a}$. The spin-orbit coupling parameter values are varied in different panels. In (a1) we set $\alpha=1,\beta=1$, (a2) $\alpha=1,\beta=-1$, (b1) $\alpha=1,\beta=1$, (b2) $\alpha=1,\beta=-1$, and in (b3) $\alpha=1,\beta=0.5$ is considered.}
\end{SCfigure*}

\subsection{Ballistic regime: Bogoliubov de Gennes formalism}\label{sec.theory.ballistic}

Consider a situation where a system is able to develop superconductivity through the traditional opposite-spin electron-phonon BCS mechanism. To describe electronic characteristics, one can use spin-Nambu field operators and introduce phonon mediated electron-electron amplitudes:
\begin{equation}
\Delta \langle \psi_\uparrow^\dag \psi_\downarrow^\dag \rangle + \text{H.c.}~.
\end{equation}  
The low-energy Hamiltonian, governing electron-hole behavior in the BCS superconductivity, reads
\begin{equation}\label{Hamil}
{\cal H}(\mathbf{p}) = \left( \begin{array}{cc}
 H(\mathbf{p}) -\mu \hat{1}& \hat{\Delta} \\
 \hat{\Delta}^\dag & -H^\dag(-\mathbf{p}) +\mu \hat{1}
\end{array}\right),
\end{equation}
in which $\mu$ is the chemical potential multiplied by $2$$\times $$2$ unity matrix $\hat{1}$ and $\hat{\Delta} $ is a $2$$\times $$2$ superconducting gap matrix in spin space. Here $H(\mathbf{p})$ can be obtained by setting $\boldsymbol{\eta} _\text{so}=\mathbf{h}=0$ in Eq. (\ref{Hamil_so}). The field operators in the rotated particle-hole and spin basis are given by $\hat{\psi}=(\psi_{\uparrow}, \psi_{\downarrow}, \psi_{\downarrow}^{\dag}, -\psi_{\uparrow}^{\dag})^{\mathrm{T}}$. In the actual calculation of supercurrent across the Josephson configuration shown in Fig. \ref{fig1}, $\Delta$ is assumed nonzero in the superconducting electrodes (regions 1 and 3) and zero otherwise (region 2). Nonetheless, the superconductivity can leak into the nonsuperconducting region by the virtue of proximity effect and consequently, a location-dependent minigap can exist in the region 2 of Fig. \ref{fig1}.

One of the most important experimentally observable quantities is the current due to moving charged particles. To calculate the charge current, flowing through a two-dimensional system in the presence of SOI, magnetization, and superconductivity, we switch to real space: $\mathbf{ r}\equiv(x,y,0)$. Also, a situation with no charge sink or source is considered. In this case, the time variation of charge density vanishes, $\partial_t\rho_\text{c}\equiv 0$, in its quantum mechanical definition:
\begin{equation}\label{crntdif}
\begin{split}
\frac{\partial \rho_\text{c}}{\partial t}=\lim\limits_{\mathbf{r}\rightarrow \mathbf{r}'}\sum\limits_{\sigma\tau\sigma'\tau'}\frac{1}{i}\Big[ \psi^\dag_{\sigma\tau}(\mathbf{r}'){\cal H}_{\sigma\tau\sigma'\tau'}(\mathbf{r})\psi_{\sigma'\tau'}(\mathbf{r})\\-\psi^\dag_{\sigma\tau}(\mathbf{r}'){\cal H}_{\sigma\tau\sigma'\tau'}^\dag(\mathbf{r}')\psi_{\sigma'\tau'}(\mathbf{r})\Big].
\end{split}
\end{equation}
Here ${\cal H}_{\sigma\tau\sigma'\tau'}$ is the component form of Eq.~(\ref{Hamil}) and $\sigma, \tau$ indices label the spin and particle-hole degrees of freedom, respectively. 
Incorporating the current conservation law, the charge current density reads,
\begin{align}
\mathbf{ J}_\text{c} =\int \hspace{-.1cm} d\mathbf{r}\Big\{\hat{\psi}^\dagger(\mathbf{r}) \overrightarrow{{\cal H}}(\mathbf{r})\hat{\psi}(\mathbf{r})-
\hat{\psi}^\dagger(\mathbf{r}) \overleftarrow{{\cal H}}(\mathbf{r})\hat{\psi}(\mathbf{r}) \Big\},
\end{align}
where ${\cal H}(\mathbf{r})$ is given by Eq.~(\ref{Hamil}), after the substitution $\mathbf{ p}\equiv -i m^{-1} (\partial_x,\partial_y,0)$. The arrow directions indicate the specific wavefunctions that the Hamiltonian acts on. All the electronic and geometrical properties of a system are indeed encoded into the Hamiltonian ${\cal H}(\mathbf{r})$, its associated wavefunctions, and boundary conditions.

\subsection{Josephson junction set-up}\label{sec.theory.setup}

Figure \ref{fig1} displays the two-dimensional Josephson junction considered in this paper. The two-dimensional junction resides in the $xy$ plane and has finite-sized width $W$ and thickness $d$. The interface of superconductor-nonsuperconductor junctions extends along the $y$ direction at $x=0,d$. The magnetization possesses an arbitrary orientation and can be described by three components $\mathbf{h}=(h_x,h_y,h_z)$. The superconductor leads, regions $1,3$, support an externally controllable macroscopic phase difference $\varphi=\varphi_1-\varphi_3$. The phase difference can be tuned by for instance passing a magnetic field through an exterior SQUID-like geometry interconnected via junction Fig. \ref{fig1}.  

\section{Results and discussions}\label{sec.results}

This section is divided into two subsections. In Sec. \ref{subsec.ballistic.results}, we present our study of current-phase relation (CPR), $\varphi_0$-state, and in Sec. \ref{subsec.criticalcurrent} the critical supercurrent .

\subsection{Current-phase relations}\label{subsec.ballistic.results}
In Secs. \ref{sec.eta_a} and \ref{sec.eta_b}, the results of numerical analysis for the profile of CPR and self-biased current shall be given, considering two differing SOIs (\ref{so}) and two perpendicular directions to the orientation of Josephson junctions in the $xy$ plane.

\subsubsection{Current-phase relation in ${\eta}_\text{so}^\text{a}$-junction}\label{sec.eta_a}

We now proceed to study the supercurrent in the ballistic regime of a two-dimensional magnetized Josephson junction with Rashba-Dresselhaus SOI depicted in Fig. \ref{fig1}. Our numerical approach in the ballistic regime is able to simulate a situation where superconductivity, magnetism, and SOI coexist simultaneously with different parameter sets in the three regions $1,2,3$ of Fig. \ref{fig1} \cite{AlidoustWS1,AlidoustWS2,AlidoustBP1,AlidoustBP2}. Also, it accommodates cubic and higher order SOI terms, e.g., $\gamma(p_xp_y^2\sigma_x - p_yp_x^2\sigma_y) $ that might be relevant in some materials due to bulk inversion asymmetry. Nevertheless, in this paper, we focus on the linearized SOI models, given by Eqs. (\ref{so}), and restrict the presence of magnetization and SOI to the region 2 of Fig. \ref{fig1}. To obtain the supercurrent, we compute the current density perpendicular to the interfaces, e.g., $J_x$, and integrate over the junction cross section in the $y$ direction: $I(\varphi)=J_0\int_{-W/2}^{+W/2}  dy J_x(x,y,\varphi)$ with $J_0=2|e| |\Delta| /\hbar$, in which $|e|$ is the electric charge unit.

Diagonalizing ${\cal H}(\textbf{p})$, Eq. (\ref{Hamil}), we obtain electronic wavefunctions $\hat{\psi}_{1,2,3}(\textbf{p})$ within the regions $1,2,3$ independently. Next, the wavefunctions are matched at the left $\hat{\psi}_1$=$\hat{\psi}_2|_{x=0}$ and the right boundaries $\hat{\psi}_2$=$\hat{\psi}_3|_{x=d}$, and also the continuity condition $(\partial_\textbf{p} {\cal H}_1)_{\textbf{r}}\hat{\psi}_1$=$(\partial_\textbf{p} {\cal H}_2)_{\textbf{r}}\hat{\psi}_2|_{x=0}$, $(\partial_\textbf{p} {\cal H}_2)_{\textbf{r}}\hat{\psi}_2$=$(\partial_\textbf{p} {\cal H}_3)_{\textbf{r}}\hat{\psi}_3|_{x=d}$ is applied at these intersections. The index $\textbf{r}$ indicates that we switch to real space after taking the derivatives in momentum space. It is worth mentioning that we shall apply no simplifying assumptions and approximations to the wavefunctions in actual numerical calculations. This, however, results in highly complicated and lengthy expressions for the wavefunctions and supercurrent. Thus, we are able to evaluate them numerically only and omit giving explicit expressions. In the numerics, we consider a rather thin and wide junction $W/d\gg 1$ to avoid any finite size effect induced by the edges at $y=\pm W/2$. 

\begin{figure*}
\centering
\includegraphics[width=18.0cm,height=9.2cm]{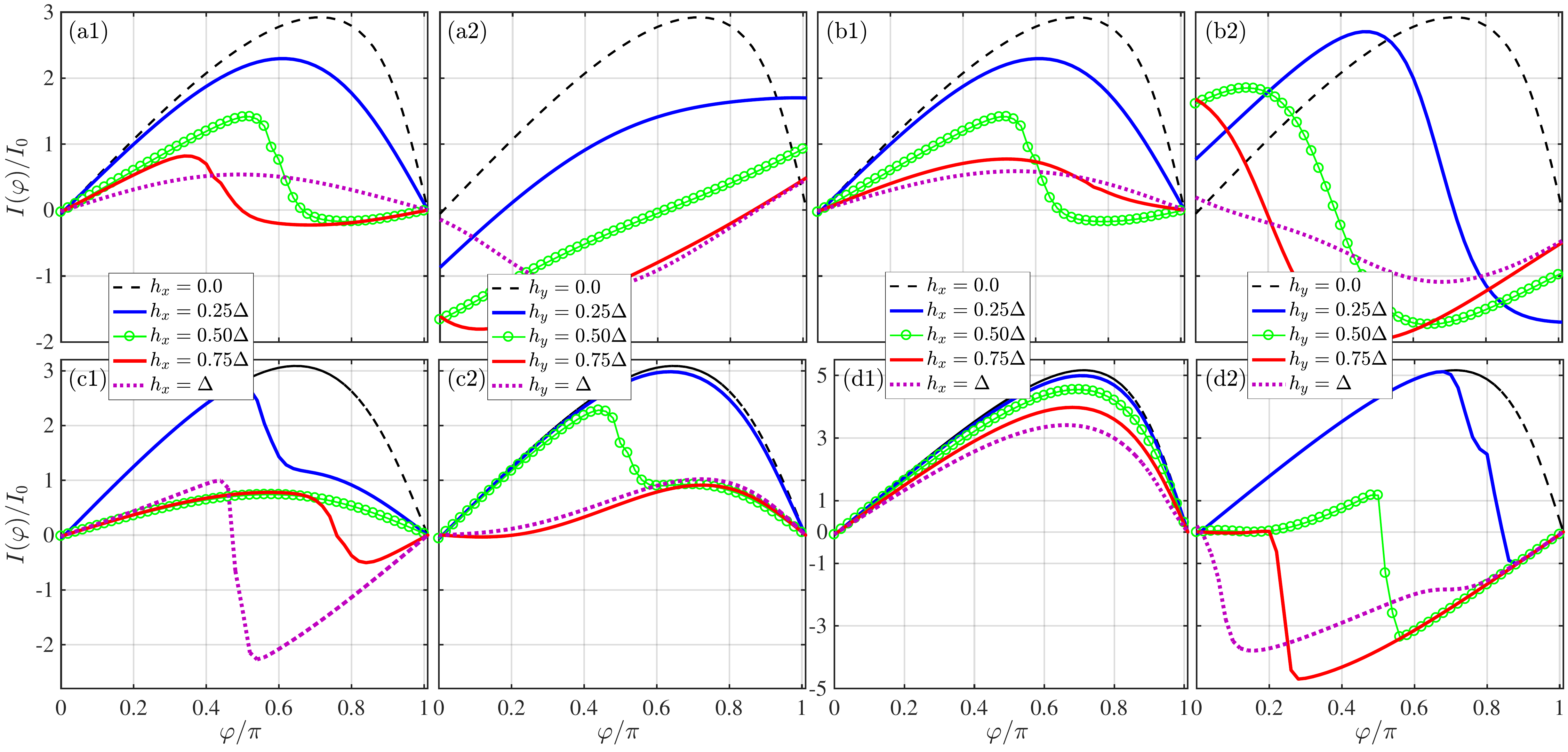}
\caption{\label{fig4} (Color online). Supercurrent vs superconducting phase difference in a Josephson junction along the $x$ direction with $\boldsymbol{\eta}_\text{so}^\text{b}$ spin-orbit coupling. The coefficients of spin-orbit interaction are: in (a1)-(a2) $\alpha=1,\beta=0$, (b1)-(b2) $\alpha=0,\beta=1$, 
(c1)-(c2) $\alpha=1,\beta=1$, and (d1)-(d2) $\alpha=1,\beta=-1$. The magnetization is oriented along the $x$ direction, i.e., $h_y=h_z=0, h_x\neq 0$, in the first and third columns, while $h_x=h_z=0, h_y\neq 0$ is set in the second and fourth columns.
}
\end{figure*}

To explore different aspects of how the interplay of RDSOI and magnetization orientation influences supercurrent, we conduct a systematic numerical study. Our extensive investigation demonstrated that supercurrent responds relatively weak to the interplay of $h_z$ magnetization and RDSOIs Eqs. (\ref{so}) ($h_z$ induces supercurrent reversal only). In addition, the interaction of $h_z$ with $\boldsymbol{\eta}_\text{so}^\text{a,b}$ is unable to induce $\varphi_0$ self-biased supercurrent. Therefore, in the following, we concentrate on supercurrent response to the interplay of in-plane magnetization $\mathbf{h}=(h_x,h_y,0)$ with RDSOC. In the numerics that follow, we consider representative values $0,\pm 1$ for the coefficients of RDSOIs $\alpha$ and $\beta$, a low value to the chemical potential $\mu=\Delta$, and $h_x,h_y=0, \pm 0.75\Delta$ for the magnetization components unless otherwise stated. The junction thickness and width are normalized to $\xi_S=\sqrt{\hbar^2 /2m\Delta}$ and fixed values $d=0.3\xi_S$ and $W=10\xi_S$ are considered in the numerics. Nevertheless, we emphasize that our conclusions made are independent of these representative parameter values.

In what follows, we visualize a few samples of the comprehensive investigation of current-phase relation performed, to make the flow of discussions smoother, illustrate how CPRs are systematically obtained, and how $\varphi_0$-state is analyzed. Figures \ref{fig2} and \ref{fig3} show the supercurrent flow through the Josephson junction shown in Fig. \ref{fig1} where the coupling of spin and orbital degrees of freedom is modeled by Eq. (\ref{so1}). In Figs. \ref{fig2}(a1) and \ref{fig2}(a2), we consider $\alpha=1, \beta=0$. As seen, $h_x=0, h_y\neq 0$ induces a nonzero supercurrent at zero phase difference $\varphi=0$, Fig. \ref{fig2}(a2). Changing the magnetization direction to $h_y=0, h_x\neq 0$ as in Fig. \ref{fig2}(a1), the $\varphi_0$ state disappears and by increasing the strength of magnetization, the supercurrent experiences $0$-$\pi$ transition and thus reverses direction. If we consider $\alpha=0, \beta=1$, shown in Figs. \ref{fig2}(b1) and \ref{fig2}(b2), the magnetization in the $y$ direction, $h_y$, is unable to generate the $\varphi_0$ state anymore. Rather, a magnetization in the $x$ direction, $h_x$, produces $\varphi_0$-state. In Figs. \ref{fig2}(c1)-\ref{fig2}(d2), both components of RDSOI are nonzero: $\alpha=\pm\beta=1$. In this case, a nonzero in-plane magnetization suffices to generate a supercurrent at zero phase difference. As seen, this is more pronounced at $\mathbf{h}=0.75\Delta$, and $\Delta$. Another interesting parameter set includes a magnetization with nonzero $h_x,h_y$. Figure \ref{fig3} exhibits the profile of CPR, $I(\varphi)$, when the in-plane magnetization is described by two components $h_x$ and $h_y$. In Figs. \ref{fig3}(a1) and \ref{fig3}(a2), the coefficients of Rashba and Dresselhaus spin-orbit coupling have equal and opposite signs: $\beta=\alpha=1$ and $\alpha=-\beta=1$, respectively, and $h_x=h_y$ is considered. Comparing with Figs. \ref{fig2}, the supercurrent at zero phase difference, i.e., the $\varphi_0$ state, vanishes for all values of the magnetization strength. The only difference between Figs. \ref{fig3}(a1) and \ref{fig3}(a2) is the $0$-$\pi$ transition response of supercurrent to the magnetization strength, which is absent when $\beta=\alpha=1$. The supercurrent monotonically decreases with increasing the magnetization strength, similarly to the supercurrent response to the junction thickness in a conventional SNS junction or a magnetic SFS junction where F is sandwiched between two ferromagnetic layers with a conical magnetization pattern \cite{jap_al}. In Figs. \ref{fig3}(b1)-\ref{fig3}(b3), we examine the influence of unequal orientation and strength in $h_x$ and $h_y$. Figure \ref{fig3}(b1) illustrates that a sign change in $h_y$, when $\beta=\alpha=1$, causes supercurrent reversal, still with no $\varphi_0$-state. The same phenomenon occurs in $\alpha=-\beta=1$ regime by changing the sign of either $h_x$ or $h_y$. Introducing inequality in the magnitude of SOCs ($|\alpha|\neq |\beta|$) or magnetization components ($|h_x|\neq |h_y|$) generates the self-biased supercurrent. This is apparent in Figs. \ref{fig3}(b1)-\ref{fig3}(b3).
We have performed an exhaustive numerical study by plotting the CPR, similar to those presented in Figs. \ref{fig2} and \ref{fig3}, for numerous sets of parameter values (not shown here). The resultant conclusions for the CPR are summarized in Eqs. (\ref{I_set_a1})-(\ref{I_set_a5}). We have shown the status of the $\varphi_0$ state in front of each set. The phase shift carries ``a,b'' labels for the spin-orbit interaction type, Eqs. (\ref{so1}) and (\ref{so2}), and ``$x,y$'' labels for the orientation of Josephson junction:

\begin{figure*}
\centering
\includegraphics[width=18.0cm,height=9.2cm]{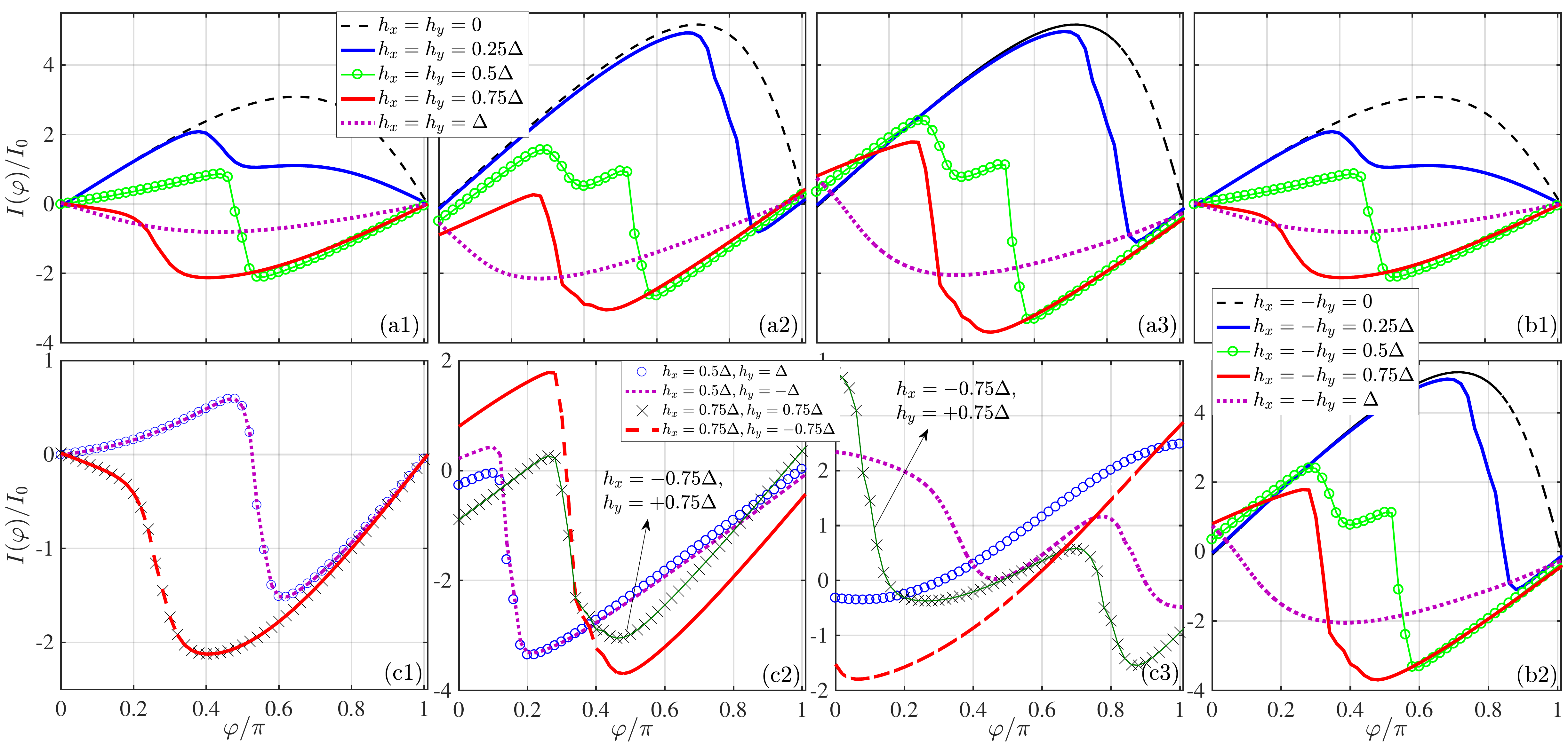}
\caption{\label{fig5} (Color online). Supercurrent vs superconducting phase difference in a $\boldsymbol{\eta}_\text{so}^\text{b}$ spin-orbit-coupled Josephson junction oriented along the $x$ direction. The magnetization has nonzero in-plane components: $h_x\neq 0, h_y\neq 0, h_z=0$. The coefficients of spin-orbit interaction vary in different panels: (a1) $\alpha=1,\beta=1$, (a2) $\alpha=1,\beta=-1$, (a3) $\alpha=-1,\beta=1$, (b1) $\alpha=1,\beta=1$, (b2) $\alpha=1,\beta=-1$, (c1) $\alpha=1,\beta=1$, (c2) $\alpha=1,\beta=-1$, and (c3) $\alpha=1,\beta=0.5$.
}
\end{figure*}

\begin{subequations}\label{I_set_a1}
\begin{eqnarray}
&& I(\pm\alpha,0,0,\mp h_y)=I(0,\pm\beta, \pm h_x, 0);\; \varphi_0^{\text{a},x}>0,~~~~\\
&& I(\pm\alpha,0,0,\pm h_y)=I(0,\pm\beta, \mp h_x, 0);\; \varphi_0^{\text{a},x}<0,~~~~
\end{eqnarray}
\end{subequations}
\begin{eqnarray}\label{I_set_a2}
&& I(\pm\alpha,0,\pm h_x,0)=I(\pm\alpha,0,\mp h_x,0)=\nonumber\\&&
I(0,\pm\beta,0,\pm h_y)=I(0,\pm \beta,0,\mp h_y);\; \varphi_0^{\text{a},x}=0,~~~~
\end{eqnarray}
\begin{subequations}\label{I_set_a3}
\begin{eqnarray}
&& I(\pm\alpha,\pm\beta, \pm h_x, 0)=I(\pm\alpha,\mp\beta, \mp h_x, 0)=\nonumber\\&&
I(\pm\alpha,\mp\beta, 0, \mp h_y)=I(\pm\alpha,\pm\beta, 0, \mp h_y);\; \varphi_0^{\text{a},x}>0,~~~~~\\&& 
I(\pm\alpha,\pm\beta, \mp h_x, 0)=I(\pm\alpha,\mp\beta, \pm h_x, 0)=\nonumber\\&&
I(\pm\alpha,\pm\beta, 0, \pm h_y)=I(\pm\alpha,\mp\beta, 0, \pm h_y); \; \varphi_0^{\text{a},x}<0,~~~~~
\end{eqnarray}
\end{subequations}
\begin{subequations}\label{I_set_a4}
\begin{align}
& I(\pm\alpha,\pm\beta, \pm h_x, \pm h_y)=I(\pm\alpha,\pm\beta, \mp h_x, \mp h_y)=\nonumber\\&
I(\pm\alpha,\mp\beta, \pm h_x, \mp h_y)=I(\pm\alpha,\mp\beta, \mp h_x, \pm h_y);\; \varphi_0^{\text{a},x}=0,~~~~~\\
& I(\pm\alpha,\mp\beta, \pm h_x, \pm h_y)=I(\pm\alpha,\mp\beta, \mp h_x, \mp h_y)=\nonumber\\& I(\pm\alpha,\pm\beta, \pm h_x, \mp h_y)=I(\pm\alpha,\pm\beta, \mp h_x, \pm h_y);\; \varphi_0^{\text{a},x}=0,~~~~~
\end{align}
\end{subequations}
\begin{subequations}\label{I_set_a5}
\begin{eqnarray}
&& I(\pm\alpha,0, +h_x,\mp h_y)=I(\pm\alpha,0, -h_x,\mp h_y)=\nonumber\\ &&
I(0,\pm\beta, \mp h_x, + h_y)=I(0,\pm\beta, \mp h_x, - h_y);\; \varphi_0^{\text{a},x}>0,~~~~~\\
&& I(\pm\alpha,0,+h_x,\mp h_y)=I(\pm\alpha,0,-h_x,\mp h_y)=\nonumber\\ &&
I(0,\pm\beta, \pm h_x, + h_y)=I(0,\pm\beta, \pm h_x, - h_y);\; \varphi_0^{\text{a},x}<0,~~~~~
\end{eqnarray}
\end{subequations}

To analyze the CPRs obtained, Eqs. (\ref{I_set_a1})-(\ref{I_set_a5}), we denote odd functions of a variable $X$ by ${\cal O}(\pm X)=\pm {\cal O}(X)$. Note that none of the odd functions ${\cal O}(X)$ in the analysis below are equal. We also consider the first-order terms for the odd functions and define an auxiliary function
\begin{eqnarray}
\Theta(X) = \theta(+X)+\theta(-X)=\left\{\begin{array}{cc}
0 & X=0\\
1 & X\neq 0
\end{array}\right. ,
\end{eqnarray}
in which $\theta(X)$ is the conventional step function. The relations Eqs.(\ref{I_set_a1}) imply that the phase shift should have a form of $\varphi_0^{\text{a},x}(0,\beta,h_x,0)\propto {\cal O}(h_x){\cal O}(\beta)$ and $\varphi_0^{\text{a},x}(\alpha,0,0,h_y)\propto -{\cal O}(h_y){\cal O}(\alpha)$. Considering Eqs. (\ref{I_set_a2}), we reaffirm that the phase shift in the absence of $h_x$ and $h_y$ is insensitive to the signs of $\beta$ and $\alpha$, respectively. One possible conclusion is therefore $\varphi_0^{\text{a},x}(\alpha,\beta,h_x,h_y)\propto {\cal O}(h_x)\beta-{\cal O}(h_y)\alpha$. This conclusion above was made when either RSOC or DSOC is available and only one component of magnetization is nonzero, i.e., Eqs. (\ref{I_set_a1}) and (\ref{I_set_a2}). Next, we keep both components of SOC nonzero and examine the influence of magnetization components separately. We have found relations Eqs. (\ref{I_set_a3}), illustrating that the above conclusion for $\varphi_0^{\text{a},x}(\alpha,\beta,h_x,h_y)$ is applicable to this parameter set as well. The relations given in Eqs. (\ref{I_set_a4}) are obtained when all components of magnetization ($h_x,h_y$) and SOC ($\alpha,\beta$) are nonzero. In all cases, the phase shift vanishes. Thus, one can conclude a phase shift of type $\varphi_0^{\text{a},x}(\alpha,\beta,h_x,h_y)\propto (\Theta(h_x)\beta^2-\Theta(h_y)\alpha^2)(h_y\alpha+h_x\beta)$. However, Eqs. (\ref{I_set_a4}) illustrate that at least a term dependent on $h_xh_y$ should be included. Hence, we examine the CPR by setting one component of SOC zero and both components of the magnetization nonzero. The results are summarized in Eqs. (\ref{I_set_a5}). Our analysis of Eqs. (\ref{I_set_a5}) together with those discussed above through Eqs. (\ref{I_set_a1})-(\ref{I_set_a4}) suggests a phase shift $\varphi_0^{\text{a},x}(\alpha,\beta,h_x,h_y)$ of the following form when spin orbital coupling $\boldsymbol{\eta}_\text{so}^\text{a}$ interacts with an in-plane magnetization:
\begin{align}\label{phi0_a_x}
\varphi_0^{\text{a},x}\propto +\Big( \Gamma_x h_y\alpha + \Gamma_yh_x\beta \Big)\Big( \Theta(h_y)\alpha^2-\Theta(h_x)\beta^2\Big).
\end{align}
Here, we have defined $\Gamma_{x,y}=\gamma \Theta(h_{x,y})- 1$ with $\gamma>1$. Note that Eq. (\ref{phi0_a_x}) can be considered as an effective phase shift to a sinusoidal CPR, namely, $\sin (\varphi+\varphi_0^{\text{a},x})$. The numerical study only allows for obtaining the functionality of phase shift with respect to different parameters. However, as is clear, the phase shift has to be a dimensionless variable and the numerical analysis is unable to provide an exact coefficient. Therefore, we keep the proportional sign ($\propto$) in the presentation of our results.

For completeness, we have performed the same numerical study as described above for a Josephson junction oriented along the $y$ axis. According to Fig. \ref{fig1}, we now solely rotate the coordinate axes by the amount of $90^\circ$ either clockwise or counter clockwise around the $z$-axis. The resultant CPRs are presented in Appendix \ref{sec.apnx.ballistic.current}, namely, Eqs.~(\ref{I_set_a1_appx})-(\ref{I_set_a5_appx}). Analyzing the functionality of Eqs.~(\ref{I_set_a1_appx})-(\ref{I_set_a5_appx}) with respect to $\alpha$, $\beta$, $h_x$, $h_y$, we have obtained the following expression for the anomalous phase shift $\varphi_0^{\text{a},y}(\alpha,\beta,h_x,h_y)$:  
\begin{align}\label{phi0_a_y}
\varphi_0^{\text{a},y}\propto -\Big( \Gamma_x h_y\beta+\Gamma_y h_x\alpha\Big)\Big( \Theta(h_y)\alpha^2-\Theta(h_x)\beta^2\Big),
\end{align}
Comparing phase shift $\varphi_0^{\text{a},x}$, Eq. (\ref{phi0_a_x}), to $\varphi_0^{\text{a},y}$, Eq. (\ref{phi0_a_y}), we find that $\varphi_0^{\text{a},y}(\alpha,\beta,h_x,h_y)=\varphi_0^{\text{a},x}(\beta,\alpha,h_x,h_y)$. This finding can be directly confirmed by the actual RDSOIs, Eqs. (\ref{so1}) and (\ref{so2}), considered in the numerics.

Another limit of interest emerges when the chemical potential is larger enough than all energies available in the system (i.e., $\mu\gg$ $\textbf{h}$, $\Delta$). To evaluate this limit in the ballistic regime, we have set $\mu=10\Delta$ and repeated the above numerical study. The resultant CPRs are similar to Eqs. (\ref{I_set_a2}), (\ref{I_set_a4}), and (\ref{I_set_a5}) except now $\varphi_0^{\text{a},x}$ reverses sign in Eqs. (\ref{I_set_a1}) and vanishes in Eqs. (\ref{I_set_a3}). By performing the analysis, the same as what is described above, for this new set of numerical CPRs we find 
\begin{align}\label{phi0_a_x_largeMu}
\varphi_0^{\text{a},x}\propto +(  h_y\alpha + h_x\beta )( \alpha^2-\beta^2),
\end{align}
and 
\begin{align}\label{phi0_a_y_largeMu}
\varphi_0^{\text{a},y}\propto -(  h_y\beta+ h_x\alpha)( \alpha^2-\beta^2).
\end{align} 
As seen, the phase-shift functionalities now reduce to relatively simpler expressions in this limit.

\subsubsection{Current-phase relation in ${\eta}_\text{so}^\text{b}$-junction}\label{sec.eta_b}

Next, we consider a Josephson junction oriented along the $x$ axis, hosting RDSOI of type $\boldsymbol{\eta}_\text{so}^\text{b}$, Eq. (\ref{so2}). Some representative cases are shown in Figs. \ref{fig4} and \ref{fig5}. In Figs. \ref{fig4}(a1) and \ref{fig4}(a2), $\alpha=1, \beta=0$ are set as the coefficients of SOC, whereas $\alpha=0, \beta=1$ are considered in Figs. \ref{fig4}(b1) and \ref{fig4}(b2). The magnetization in the first/third and second/fourth columns is oriented along the $x$ and $y$ axis, respectively. It is apparent that for both parameter sets of SOC the self-biased supercurrent appears only when the magnetization is directed along the $y$ axis. Increasing the strength of magnetization oriented along the $x$ axis, the supercurrent experiences reversal and the contribution of higher order harmonics, such as $\sin 2\varphi, \sin 3\varphi, ...$, into the supercurrent becomes more pronounced. In Figs.\ref{fig4}(c1)-\ref{fig4}(c2) and Figs.\ref{fig4}(d1)-\ref{fig4}(d2), the coefficients of SOC are $\alpha=\beta=1$ and $\alpha=-\beta=1$, respectively. As seen, the phase shift vanishes independent of magnetization orientation. The prominent difference appears in the supercurrent reversal that occurs when only $h_x$ ($h_y$) is nonzero in former (latter) set of SOC coefficients. In Fig. \ref{fig5}, we switch both magnetization components on. Figures \ref{fig5}(a1)-\ref{fig5}(a3) show the supercurrent-phase profile when magnetization components are identical $h_x=h_y$ and the spin-orbit coupling coefficients are $\alpha=\beta=1$, $\alpha=-\beta=1$, and $-\alpha=\beta=1$, respectively. When $\alpha$ and $\beta$ are positive, Fig. \ref{fig5}(a1) illustrates a zero self-biased current. Tuning them to obtain opposite signs in Figs. \ref{fig5}(a1)-\ref{fig5}(a2), the self-biased current across the junction switches direction. In Figs. \ref{fig5}(b1)-\ref{fig5}(b2), we set $\alpha=\beta=1$ and $\alpha=-\beta=1$, respectively, and opposite signs for the magnetization components $h_x=-h_y$. The figures demonstrate that the phase shift remains zero when $\alpha=\beta=1$, independent of the magnetization orientation. Finally, in Figs. \ref{fig5}(c1)-\ref{fig5}(c3) we set unequal values for both the magnetization and SOC components. The results imply that the phase shift is insensitive to the magnetization component along the $x$ axis. To shed light on the functionality and dependency of the self-biased current on $\alpha$, $\beta$, $h_x$, and $h_y$, we have performed the same systematic numerical study on the current-phase profile as described in Sec. \ref{sec.eta_a}. The results are summarized in Eqs. (\ref{I_set_b1})-(\ref{I_set_b5}):  
\begin{figure*}
\centering
\includegraphics[width=18.0cm,height=7.7cm]{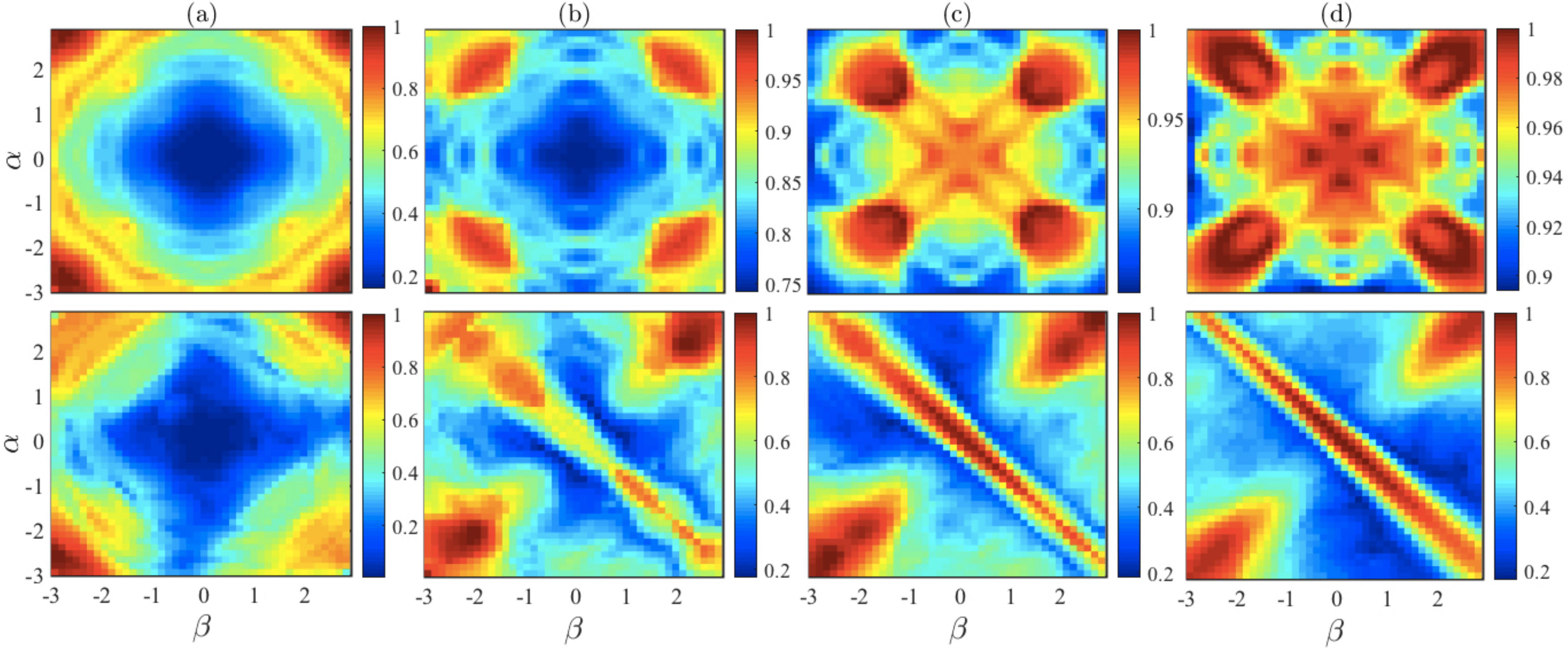}
\caption{\label{fig6} (Color online). Normalized critical supercurrent as a function of $\alpha$ and $\beta$ parameters of RDSOI, flowing in a ballistic Josephson junction. In the top row, we set $h_x=h_y=0$ while in the bottom row $h_x=h_y=0.75\Delta$. The chemical potential varies column-wise: (a) $\mu=\Delta$, (b) $\mu=5\Delta$, (c) $\mu=10\Delta$, and (d) $\mu=20\Delta$. The type of RDSOI is $\boldsymbol{\eta}^\text{a}_\text{so}$.}
\end{figure*}

\begin{subequations}\label{I_set_b1}
\begin{eqnarray}
&& I(\pm\alpha,0, 0, \mp h_y)=I(0,\pm\beta, 0, \pm h_y);\; \varphi_0^{\text{b},x}>0,~~~~~\\
&& I(\pm \alpha,0, 0, \pm h_y)=I(0,\pm \beta, 0, \mp h_y);\; \varphi_0^{\text{b},x}<0,~~~~~
\end{eqnarray}
\end{subequations}
\begin{subequations}\label{I_set_b2}
\begin{eqnarray}
&& I(\pm\alpha,\pm\beta, 0, +h_y)=I(\pm\alpha,\pm\beta, 0, -h_y);\; \varphi_0^{\text{b},x}=0,~~~~~\\
&& I(\pm \alpha,\mp\beta, 0, +h_y)=I(\pm \alpha,\mp\beta, 0, -h_y);\; \varphi_0^{\text{b},x}=0,~~~~~\\
&& I(\pm\alpha,\mp\beta, +h_x, 0)=I(\pm\alpha,\mp\beta, 0, -h_y);\; \varphi_0^{\text{b},x}=0,~~~~~\\
&& I(\pm\alpha,\pm\beta, +h_x, 0)=I(\pm\alpha,\pm\beta, -h_x, 0); \; \varphi_0^{\text{b},x}=0,~~~~~
\end{eqnarray}
\end{subequations}
\begin{subequations}\label{I_set_b3}
\begin{align}
& I(\pm\alpha,\mp\beta, \pm h_x, \mp h_y)=I(\pm\alpha,\mp\beta, \mp h_x, \mp h_y);\; \varphi_0^{\text{b},x}>0,~~~~~\\
& I(\pm\alpha,\mp\beta, \pm h_x, \pm h_y)=I(\pm\alpha,\mp\beta, \mp h_x, \pm h_y);\; \varphi_0^{\text{b},x}<0,~~~~~
\end{align}
\end{subequations}
\begin{align}\label{I_set_b4}
& I(\pm\alpha,\pm\beta, \pm h_x, \pm h_y)=I(\pm\alpha,\pm\beta, \mp h_x, \mp h_y)=\nonumber\\ & 
I(\pm\alpha,\pm\beta, \pm h_x, \mp h_y)=I(\pm\alpha,\mp\beta, \mp h_x, \pm h_y);\; \varphi_0^{\text{b},x}=0,~~~~~
\end{align}
\begin{subequations}\label{I_set_b5}
\begin{align}
& I(\pm\alpha,0,\pm h_x,\pm h_y)=I(\pm\alpha,0,\mp h_x,\pm h_y)=\nonumber\\ &
I(0,\pm\beta, \pm h_x, \mp h_y)=I(0,\pm\beta, \mp h_x, \mp h_y);\; \varphi_0^{\text{b},x}>0,~~~~~\\
& I(\pm\alpha,0,\pm h_x,\mp h_y)=I(\pm\alpha,0,\mp h_x,\mp h_y)=\nonumber\\ &
I(0,\pm \beta, \pm h_x, \pm h_y)=I(0,\pm \beta, \mp h_x, \pm h_y);\; \varphi_0^{\text{b},x}<0,~~~~~
\end{align}
\end{subequations}

Considering Eqs. (\ref{I_set_b1}), we deduce that $\varphi_0^{\text{b},x}(\alpha,\beta,h_x,h_y)$ can be expressed by $\varphi_0^{\text{b},x}(\alpha,0,0,h_y)\propto -{\cal O}(h_y){\cal O}(\alpha)$ and $\varphi_0^{\text{b},x}(0,\beta,0,h_y)\propto {\cal O}(h_y){\cal O}(\beta)$. However, in the presence of both $\alpha$ and $\beta$ with arbitrary signs, the relations Eqs. (\ref{I_set_b2}) illustrate that $\varphi_0^{\text{b},x}=0$. Thus, we can conclude $\varphi_0^{\text{b},x}(\alpha,\beta,0,h_y)\propto {\cal O}(h_y)(\beta^2-\alpha^2)(\beta+\alpha)$. Equations (\ref{I_set_b3}) and (\ref{I_set_b4}) illustrate that a term dependent on $h_xh_y$ is missing. By considering Eqs. (\ref{I_set_b1}), (\ref{I_set_b2}), (\ref{I_set_b4}), (\ref{I_set_b5}) together with Eqs. (\ref{I_set_b3}), the numerical analysis offers the following form for the anomalous phase shift
\begin{align}\label{phi0_b_x}
&\varphi_0^{\text{b},x}\propto +\Gamma_x h_y(\alpha+\beta)(\alpha^2-\beta^2).
\end{align}

We have also conducted this systematic study when the Josephson junction is oriented along the $y$ axis, supporting the $\boldsymbol{\eta}_\text{so}^\text{b}$ type of SOC. We have summarized the resultant CPRs and the status of the corresponding phase shift in Appendix \ref{sec.apnx.ballistic.current}; Eqs. (\ref{I_set_b1_appx})-(\ref{I_set_b5_appx}). The same analysis as the above yields the following relation for the phase shift in this case:  
\begin{align}\label{phi0_b_y}
&\varphi_0^{\text{b},y}\propto -\Gamma_y h_x(\alpha-\beta)(\alpha^2-\beta^2).
\end{align}
Comparing $\varphi_0^{\text{b},x}$, Eq. (\ref{phi0_b_x}), to $\varphi_0^{\text{b},y}$, Eq. (\ref{phi0_b_y}), we find $\varphi_0^{\text{b},y}(\alpha,\beta,h_y)=\varphi_0^{\text{b},x}(-\alpha,\beta,h_x)$, which is in full agreement with Eq. (\ref{so2}).

By setting a large value to the chemical potential, i.e., $\mu=10\Delta$ and numerically reproducing CPRs, we find that $\varphi_0^{\text{b},x}$ reverses sign in Eqs. (\ref{I_set_b1}) and vanishes in Eqs. (\ref{I_set_b3}). The rest of the CPRs given by Eqs. (\ref{I_set_b2}), Eqs. (\ref{I_set_b4}), and Eqs. (\ref{I_set_b5}) remain unchanged. The analysis of these new CPRs yields the following expressions for the anomalous phase shifts, when the Josephson junction is oriented along the $x$ axis:
 \begin{align}\label{phi0_b_x_LargeMu}
&\varphi_0^{\text{b},x}\propto +h_y(\alpha+\beta)(\alpha^2-\beta^2),
\end{align}
and when the junction is rotated and the total supercurrent flows along the $y$ axis:
\begin{align}\label{phi0_b_y_LargeMu}
&\varphi_0^{\text{b},y}\propto - h_x(\alpha-\beta)(\alpha^2-\beta^2).
\end{align}

Having obtained the explicit functionalities for the anomalous phase shift as a function of magnetization and SOC components, we now present the results for critical supercurrent.

\subsection{Critical supercurrent}\label{subsec.criticalcurrent}

\begin{figure*}
\centering
\includegraphics[width=18.0cm,height=7.7cm]{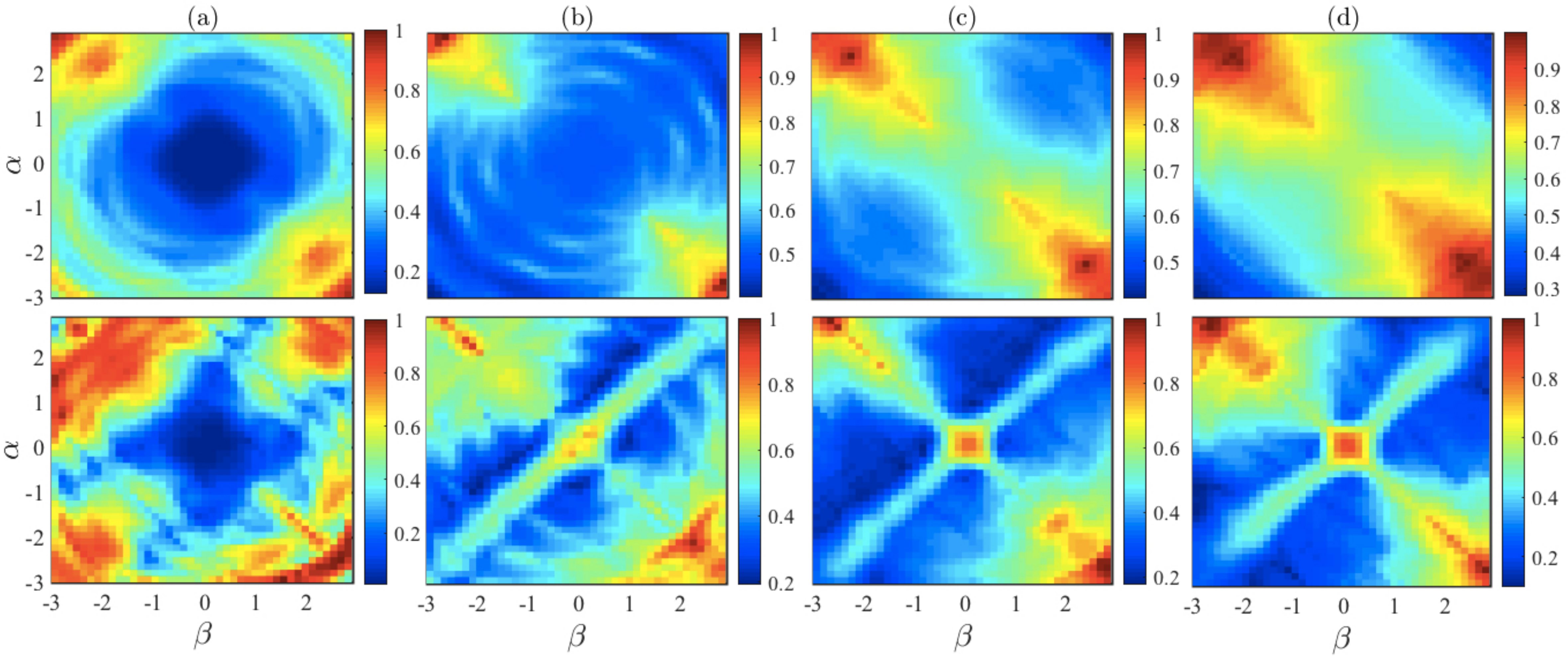}
\caption{\label{fig7} (Color online). The color-map profile of normalized maximum supercurrent in a ballistic system as a function of RDSOI parameters, $\alpha$ and $\beta$, where RDSOI is described by $\boldsymbol{\eta}_\text{so}^\text{b}$. From left to right the chemical potential increases: (a) $\mu=\Delta$, (b) $\mu=5\Delta$, (c) $\mu=10\Delta$, (d) $\mu=20\Delta$. In the top row, the Zeeman field is switched off, $h_x=h_y=0$, and in the bottom row the Zeeman field components are equal and nonzero $h_x=h_y=0.75\Delta$.}
\end{figure*}
We adopt the same assumptions made in the previous section. Namely, the superconductor leads are made of conventional superconductors, RDSOI and Zeeman field are restricted within the nonsuperconducting region, and the inverse proximity effect is negligible. For both ballistic and diffusive regimes, $W\gg d$ ($W=10d$) is considered to avoid edge dictated effects and the Zeeman field considered is in-plane, i.e., $\mathbf{h}=(h_x,h_y,0)$. The energies and lengths are normalized by the superconducting energy gap $\Delta$ and superconducting coherence length $\xi_S$, respectively. The superconducting coherence length in the diffusive regime is given by $\xi_S=\sqrt{D/\Delta}$ in which $D$ is the diffusion constant \cite{alidoust1,alidoust2}. The thickness of ballistic junction is set fixed at $d=\xi_S$ throughout the numerics. By varying the superconducting phase difference $\varphi=\varphi_{l}-\varphi_{r}$ within $[0,2\pi]$ interval, we compute the supercurrent phase relation $I(\varphi)$ and determine critical supercurrent by $I_\text{max}=\text{max}\left( |I(\varphi;\alpha,\beta,\mathbf{h})|\right)$.

Figure \ref{fig6} exhibits the critical supercurrent as a function of RDSOI parameters, i.e., $\alpha$ and $\beta$. The supercurrent in each panel is normalized by its maximum value. Here, RDSOI is described by Eq. (\ref{so1}), and the Zeeman field is zero in top row while $h_x=h_y=0.75\Delta$ is set in bottom row. The chemical potential increases from column (a) to (d); $\mu=\Delta$, $\mu=5\Delta$, $\mu=10\Delta$, and $\mu=20\Delta$, respectively. In column (a) of Fig. \ref{fig6}, a low chemical potential has washed out any prominent impacts that the persistent spin helices may have on the critical supercurrent. The addition of the Zeeman field is also unable to recover any finger prints of the persistent spin helices. Increasing the chemical potential, the signature of the persistent spin helices appears. As seen, maximal critical supercurrent tends to occur at and around $|\alpha|=|\beta|$. This is more pronounced in the top panel of Fig. \ref{fig6}(d) where $\mu=20\Delta$. The application of an in-plane Zeeman field $h_x=h_y=0.75\Delta$ in the lower panels of Figs. \ref{fig6}(b)-\ref{fig6}(d) eliminates the extra features away from $|\alpha|=|\beta|$, appearing when $h_x=h_y=0$, and clearly illustrates how the persistent spin helices enhance the critical supercurrent.

\begin{figure*}
\centering
\includegraphics[width=18.0cm,height=8.7cm]{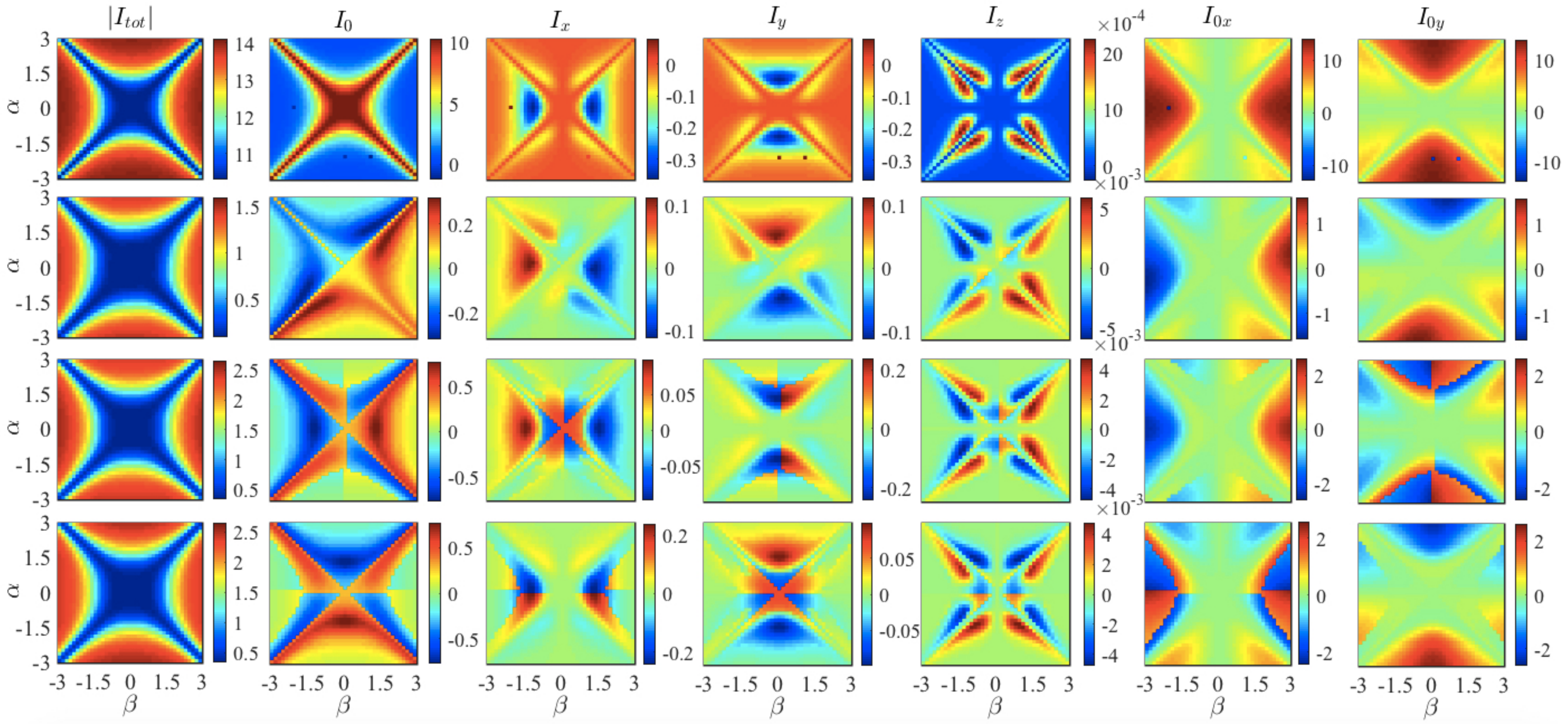}
\caption{\label{fig8} (Color online). Modulus of critical supercurrent ($|I_\text{tot}|$) and its components as a function of RDSOI parameters, $\alpha$ and $\beta$, in a diffusive system. The RDSOI is described by $\boldsymbol{\eta}_\text{so}^\text{a}$. From top to bottom: first row: $h_x=0.0, h_y=0.0$; second row: $h_x=1.5\Delta, h_y=1.5\Delta$; third row: $h_x=1.5\Delta, h_y=0.0$, and fourth row: $h_x=0.0, h_y=1.5\Delta$.}
\end{figure*}

Next, we perform the same study except we now consider $\boldsymbol{\eta}_\text{so}^\text{b}$ as RDSOI. Figure \ref{fig7} shows the critical supercurrent vs $\alpha$ and $\beta$ parameters of RDSOI. Similar to $\boldsymbol{\eta}_\text{so}^\text{a}$ RDSOI, a low chemical potential causes insignificant indication of the persistent spin helices. The increase of chemical potential results in a notable trace of persistent spin helices. Specifically, when the Zeeman field is absent in the top row of Fig. \ref{fig7}, the maximal supercurrent is localized around $|\alpha|=-|\beta|$. In the presence of the in-plane Zeeman field $h_x=h_y=0.75\Delta$, shown in the bottom row of Fig. \ref{fig7}, the maximal supercurrent passing through the junction occurs when $|\alpha|=|\beta|$ akin to $\boldsymbol{\eta}_\text{so}^\text{a}$ RDSOI. 

To illustrate how a Zeeman field also can weaken the indication of the persistent spin helices through critical supercurrent, we plot the critical supercurrent vs the $\alpha$ and $\beta$ parameters of $\boldsymbol{\eta}_\text{so}^\text{a}$ and $\boldsymbol{\eta}_\text{so}^\text{b}$ RDSOI in Figs. \ref{fig10} and \ref{fig11} of the Appendix, the two components of Zeeman field are different in magnitude ($h_x\neq h_y$). As seen, a unidirectional Zeeman field ($h_x=0, h_y=0.75\Delta$) in the presence of $\boldsymbol{\eta}_\text{so}^\text{a}$ (top row of Fig. \ref{fig10}) or $\boldsymbol{\eta}_\text{so}^\text{b}$ (bottom row of Fig. \ref{fig11}) introduces a detrimental effect on the signature of the persistent spin helices, while ($h_x=\Delta, h_y=0.75\Delta$) can still reveal a prominent indication of the persistent spin helices. Figure \ref{fig11} shows the critical current when $\boldsymbol{\eta}_\text{so}^\text{b}$ is present. Here, in the top and bottom rows, $h_x=0.75\Delta, h_y=0$ and $h_x=0, h_y=0.75\Delta$ are set, respectively. Comparing to Fig. \ref{fig7}, $h_x=0.75\Delta, h_y=0$ has eliminated the indication of the persistent spin helices stronger than $h_x=0, h_y=0.75\Delta$ case. As is well understood, the Andreev subgap states in the ballistic regime play a pivotal role in carrying supercurrent from one superconductor to another and Zeeman field strongly alters these subgap channels. Hence, various combinations of $h_x$ and $h_y$ components may be beneficial or detrimental when studying SOIs through critical supercurrent in ballistic junctions. 

\begin{figure*}
\centering
\includegraphics[width=18.0cm,height=8.7cm]{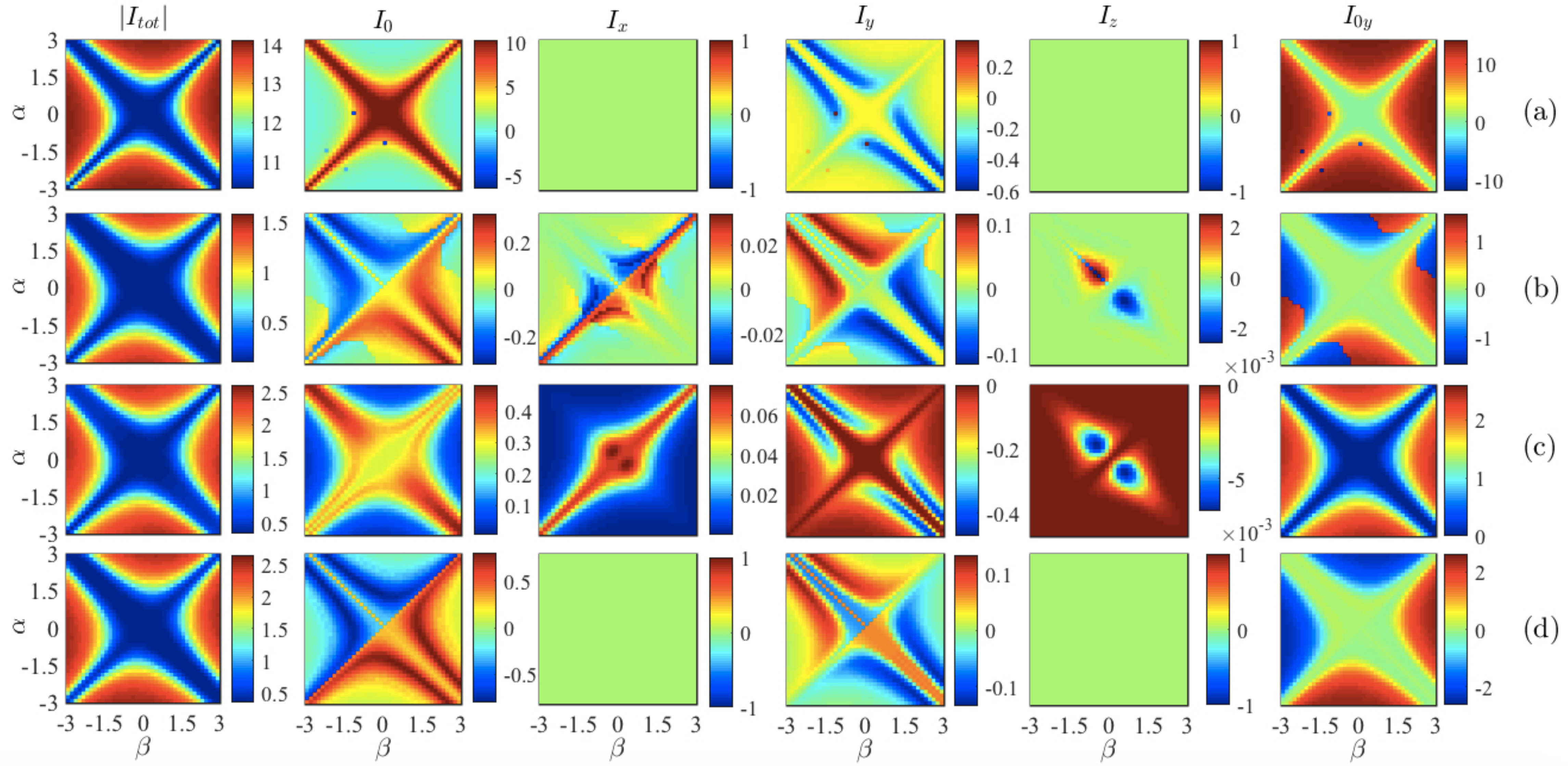}
\caption{\label{fig9} (Color online). Color-map profile of critical supercurrent modulus ($|I_\text{tot}|$) and its components vs $\alpha$ and $\beta$, the parameters of $\boldsymbol{\eta}_\text{so}^\text{b}$ RDSOI in a diffusive Josephson junction. The parameters in each row is set as follows: (a) $h_x=0.0, h_y=0.0$, (b) $h_x=1.5\Delta, h_y=1.5\Delta$, (c) $h_x=1.5\Delta, h_y=0.0$, and (d) $h_x=0.0, h_y=1.5\Delta$.}
\end{figure*}

For completeness, we turn our attention to the diffusive regime of the same Josephson configuration described earlier. Following Ref. \onlinecite{jap_al}, one can decompose the current into components, in the presence of $\boldsymbol{\eta}_\text{so}^\text{a}$ and $\boldsymbol{\eta}_\text{so}^\text{b}$ RDSOI, so that the current components contain specific components of the Green's function $(f_0,f_x,f_y,f_z)$:
\begin{align}\label{current_comp_a}
&\mathbf{J}= \sum_{i,j=0,x,y,z}   \mathbf{J}_{ij}.
\end{align}
By the above decomposition, one is able to isolate the contribution of spin-singlet $f_0$, spin triplet $f_x, f_y, f_z$ superconducting correlations, and the multiplication of spin-singlet and spin-triplet terms into a critical supercurrent flow. For example, $J_0(\equiv J_{00})$ and $J_{x,y,z}(\equiv J_{xx,yy,zz})$ can provide a rich overview for pure spin-singlet and spin-triplet supercurrents, respectively, whereas $J_{0x}$, $J_{0y}$, $J_{xz}$, $J_{yz}$ contain crossed terms, involving both the spin-singlet and spin-triplet Green's functions. To determine the critical supercurrent accurately, we solve the coupled Usadel equations together with boundary conditions numerically, following Refs. \onlinecite{alidoust1,alidoust2}. Substituting the numerical solutions into the current definition of Refs. \onlinecite{alidoust1,alidoust2}, we obtain the supercurrent as a function of phase difference $\varphi$, i.e., $I(\varphi)$, and determine the critical current: $|I_\text{tot}|=\text{max}\left( |I(\varphi; \alpha,\beta,\mathbf{h})|\right)$. The current is normalized by $\pi \sigma/e\Delta$ throughout the supercurrent calculation in the diffusive regime. Next, we find the specific phase difference that causes maximum supercurrent $\varphi_\text{max}$ and obtain the current components at $\varphi_\text{max}$. The numerical results are summarized in Figs. \ref{fig8} and \ref{fig9} where RDSOI is described by $\boldsymbol{\eta}_\text{so}^\text{a}$ and $\boldsymbol{\eta}_\text{so}^\text{b}$, respectively. The absolute value of total supercurrent is shown in the left most column, labeled by $|I_\text{tot}|$, and the associated components described above are given in the rest of the columns. The junction thickness in the diffusive regime is set fixed at $d=2\xi_S$. The critical supercurrent is plotted as a function of $\alpha$ and $\beta$ parameters for four sets of Zeeman field components from top to bottom; first row: $h_x=0, h_y=0$, second row: $h_x=1.5\Delta, h_y=1.5\Delta$, third row: $h_x=1.5\Delta, h_y=0$, and fourth row: $h_x=0, h_y=1.5\Delta$. Comparing to the ballistic Josephson junction, the signature of the persistent spin helices is drastically changed. In both cases of RDSOI, the critical supercurrent is diminished when $|\alpha|=|\beta|$ and yields a significant indication of the persistent spin helices by suppressing the critical supercurrent. The current component plots illustrate that the main contributing components of supercurrent are the spin-singlet component $I_0$ (associated with $f_0$), spin triplet components $I_x, I_y$, and crossed terms $I_{0,x}$ and $I_{0,y}$. The results reveal that the pure spin-singlet component $I_0$ reaches maximal values at and around $|\alpha|=|\beta|$, similar to the ballistic case. However, the presence of the spin-triplet components, and consequently crossed terms, dominates and enhances the supercurrent away from $|\alpha|=|\beta|$ in all cases. This phenomenon is more prominent when the junction thickness increases. Note that the presence of the Zeeman field causes weak modifications to the modulus of the supercurrent $|I_\text{tot}|$, although it can induce nonzero spin triplet supercurrent in the case of $\boldsymbol{\eta}_\text{so}^\text{b}$ (compare panels $I_x$ in Fig. \ref{fig9}). In the diffusive regime, due to strong scattering resources (disorder and nonmagnetic impurities), the spin-singlet component of supercurrent highly suppresses by increasing the junction thickness \cite{jap_al}. However, the spin-triplet components, involving $f_x$ and $f_y$, are long range, weakly sensitive to the scattering resources, and propagate over the entire nonsuperconducting region with a large decaying length scale. Therefore, in a thick enough junction, the variation of the critical supercurrent is governed by the spin triplet components.

\section{Conclusions}\label{sec.conclusion}

In summary, we have theoretically studied self-biased supercurrent (so called $\varphi_0$-Josephson state) in a two-dimensional Josephson junction driven by the interplay of Zeeman field with in-plane components ($h_x$,$h_y$,$0$) and two differing types of Rashba($\alpha$)-Dresselhaus($\beta$) spin-orbit couplings (RDSOCs). In ballistic regime, we solve the Bogoliubov de Gennes numerically without incorporating simplifying assumptions to its associated wavefunctions and study current-phase profile by employing numerous sets of parameter values. Analyzing numerical results, we obtain explicit functionalities for the $\varphi_0$ phase shift with respect to the components of magnetization and RDSOC parameters. The findings illustrate that $|\alpha|$=$|\beta|$ removes the $\varphi_0$ phase shift independent of magnetization direction and strength, the density of nonmagnetic impurities, and junction direction, when $\mu$ is high enough compared to the energy gap ($\Delta$) in the superconductor leads; $\mu\gg\Delta$. In the $\mu\sim \Delta$ limit, however, except a certain case where $|h_x|=|h_y|$, the magnetization retrieves the $\varphi_0$ phase shift. In striking contrast to the $\mu\gg \Delta$ limit, the $\varphi_0$ phase shift in the $\mu\sim \Delta$ limit is directly proportional to $h_xh_y$ terms. 
Also, we find that a low chemical potential compared to the superconducting gap $\mu\sim\Delta$ is unfavorable for detecting the persistent spin helices whereas in a ballistic and short junction with $\mu\gg\Delta$, the maximum of critical supercurrent is localized at and around $|\alpha|=|\beta|$ symmetry lines. We find that a proper fine-tunning of the in-plane Zeeman field can cause more pronounced indications for the persistent spin helices in the ballistic regime. In diffusive regime, we employ a quasiclassical technique that allows for isolating the spin-singlet and spin-triplet components of supercurrent. We show that due to the contribution of long-range spin-triplet supercurrent away from $|\alpha|=|\beta|$ symmetry lines, the critical supercurrent suppresses at and around $|\alpha|=|\beta|$. Considering the accessibility of the $\varphi_0$ phase shift in both the ballistic and diffusive systems and the tuneability of RDSOC, the uncovered $\varphi_0$ expressions and the behavior of critical supercurrent can be utilized as a tool for characterizing the type of spin-orbital interaction, confirming controllable persistent spin helices, experimentally extracting reliable values for the parameters of spin-orbital interaction in a system, and corroborating the existence of long-range spin-triplet superconducting correlations. Our results can be confirmed by $\rm GaAs$ quantum wells and zinc-blende materials.

\acknowledgments
M.A. is supported by Iran's National Elites Foundation (INEF). M.A. would like to thank A. Zyuzin for stimulating discussions.

\onecolumngrid
\appendix

\section{Supercurrent in Ballistic regime}\label{sec.apnx.ballistic.current}

The supercurrent-phase relation in the presence of $\boldsymbol{\eta}_\text{so}^\text{a}$ spin-orbit coupling, Eq. (\ref{so1}) where the Josephson junction is oriented by $90^\circ$ around the $z$ axis so the supercurrent flows along the $y$ direction:
\begin{subequations}\label{I_set_a1_appx}
\begin{eqnarray}
&& I(+\alpha,0,+h_x,0)=I(-\alpha,0,-h_x,0)=I(0,-\beta, 0, +h_y)=I(0,+\beta, 0, -h_y);\;\;\;\varphi_0^{\text{a},y}>0,\\
&& I(-\alpha,0,+h_x,0)=I(+\alpha,0,-h_x,0)=I(0,+\beta, 0, +h_y)=I(0,-\beta, 0, -h_y);\;\;\; \varphi_0^{\text{a},y}<0,
\end{eqnarray}
\end{subequations}
\begin{eqnarray}\label{I_set_a2_appx}
&& I(+\alpha,0,0,+h_y)=I(-\alpha,0,0,+h_y)=I(+\alpha,0,0,-h_y)=I(-\alpha,0,0,-h_y)=\nonumber\\
&& I(0,+\beta,+h_x,0)=I(0,-\beta,+h_x,0)=I(0,+\beta,-h_x,0)=I(0,-\beta,-h_x,0);\;\;\; \varphi_0^{\text{a},y}=0,
\end{eqnarray}

\begin{figure*}
\centering
\includegraphics[width=18.0cm,height=7.7cm]{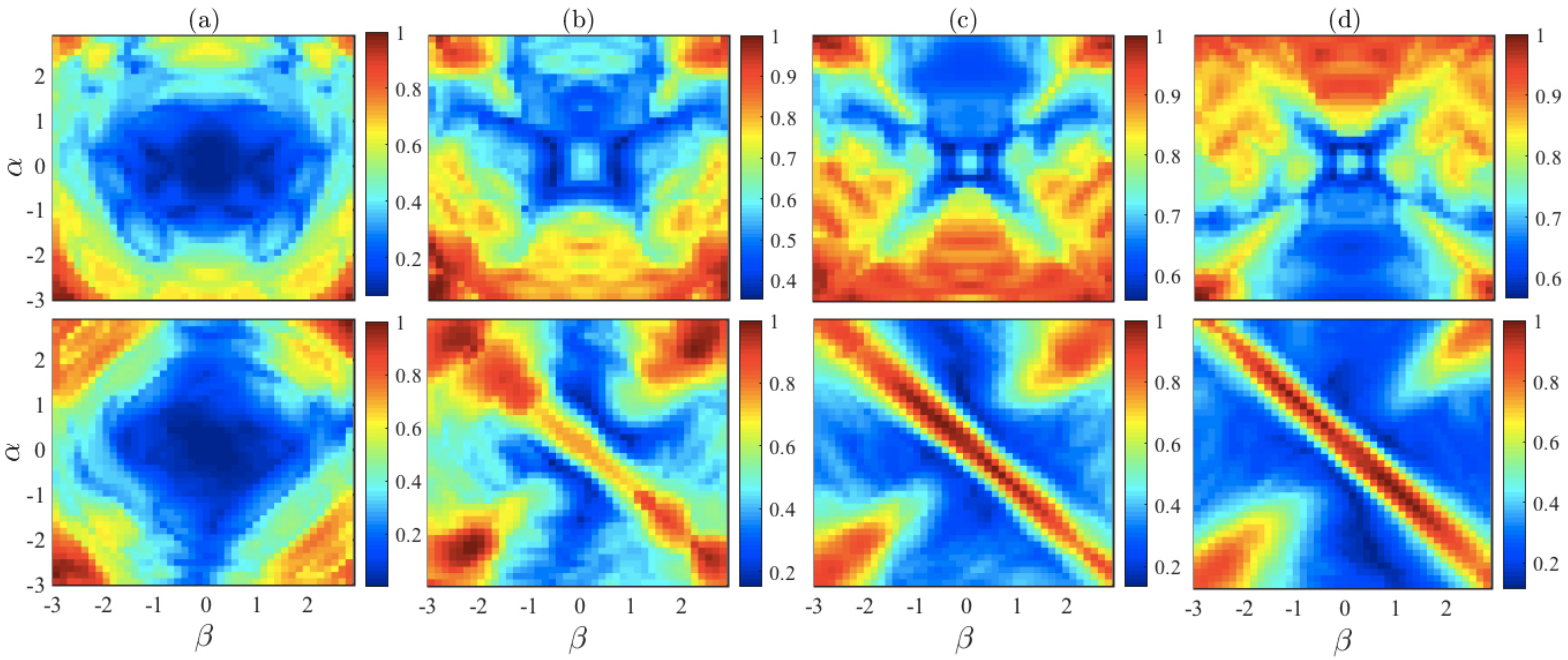}
\caption{\label{fig10} (Color online). Normalized critical current passing through a ballistic Josephson junction as a function of RDSOI parameters describing $\boldsymbol{\eta}_\text{so}^\text{a}$. In each column, a fixed value for the chemical potential is set: (a) $\mu=\Delta$, (b) $\mu=5\Delta$, (c) $\mu=10\Delta$, and (d) $\mu=20\Delta$. In the top row the components of Zeeman field are $h_x=0, h_y=0.75\Delta$, while in the bottom row $h_x=\Delta, h_y=0.75\Delta$.}
\end{figure*}

\begin{subequations}\label{I_set_a3_appx}
\begin{eqnarray}
&& I(+\alpha,+\beta, +h_x, 0)=I(+\alpha,-\beta, +h_x, 0)=I(-\alpha,+\beta, -h_x, 0)=I(-\alpha,-\beta, -h_x, 0)=\nonumber\\&& 
 I(+\alpha,-\beta, 0, +h_y)=I(-\alpha,-\beta, 0, +h_y)=I(+\alpha,+\beta, 0, -h_y)=I(-\alpha,+\beta, 0, -h_y);\;\;\ \varphi_0^{\text{a},y}>0,\\&& 
I(-\alpha,+\beta, +h_x, 0)=I(-\alpha,-\beta, +h_x, 0)=I(+\alpha,+\beta, -h_x, 0)=I(+\alpha,-\beta, -h_x, 0)=\nonumber\\&& 
 I(+\alpha,+\beta, 0, +h_y)=I(-\alpha,+\beta, 0, +h_y)=I(+\alpha,-\beta, 0, -h_y)=I(-\alpha,-\beta, 0, -h_y); \;\;\ \varphi_0^{\text{a},y}<0,
\end{eqnarray}
\end{subequations}
\begin{subequations}\label{I_set_a4_appx}
\begin{eqnarray}
&& I(+\alpha,+\beta, +h_x, +h_y)=I(-\alpha,-\beta, +h_x, +h_y)=I(+\alpha,+\beta, -h_x, -h_y)=I(-\alpha,-\beta, -h_x, -h_y)=\nonumber\\ && I(+\alpha,-\beta, +h_x, -h_y)=I(-\alpha,+\beta, +h_x, -h_y)=I(+\alpha,-\beta, -h_x, +h_y)=I(-\alpha,+\beta, -h_x, +h_y); \;\;\  \varphi_0^{\text{a},y}=0,\\
&& I(+\alpha,-\beta, +h_x, +h_y)=I(-\alpha,+\beta, +h_x, +h_y)=I(+\alpha,-\beta, -h_x, -h_y)=I(-\alpha,+\beta, -h_x, -h_y)=\nonumber\\ && I(+\alpha,+\beta, +h_x, -h_y)=I(-\alpha,-\beta, +h_x, -h_y)=I(+\alpha,+\beta, -h_x, +h_y)=I(-\alpha,-\beta, -h_x, +h_y); \;\;\  \varphi_0^{\text{a},y}=0,
\end{eqnarray}
\end{subequations}

\begin{subequations}\label{I_set_a5_appx}
\begin{eqnarray}
&& I(+\alpha,0,-h_x,+h_y)=I(+\alpha,0,-h_x,-h_y)=I(0,+\beta, +h_x, + h_y)=I(0,+\beta, -h_x, + h_y)=\nonumber\\ && I(0,-\beta, +h_x, - h_y)=I(0,-\beta, -h_x, - h_y)=I(-\alpha,0,+h_x,+h_y)=I(-\alpha,0,+h_x,-h_y); \;\;\ \varphi_0^{\text{a},y}>0,\\
&& I(+\alpha,0,+h_x,+h_y)=I(+\alpha,0,+h_x,-h_y)=I(0,+\beta, +h_x, - h_y)=I(0,+\beta, -h_x, - h_y)=\nonumber\\ && I(0,-\beta, +h_x, + h_y)=I(0,-\beta, -h_x, + h_y)=I(-\alpha,0,-h_x,+h_y)=I(-\alpha,0,-h_x,-h_y); \;\;\ \varphi_0^{\text{a},y}<0,
\end{eqnarray}
\end{subequations}

\begin{figure*}
\centering
\includegraphics[width=18.0cm,height=7.7cm]{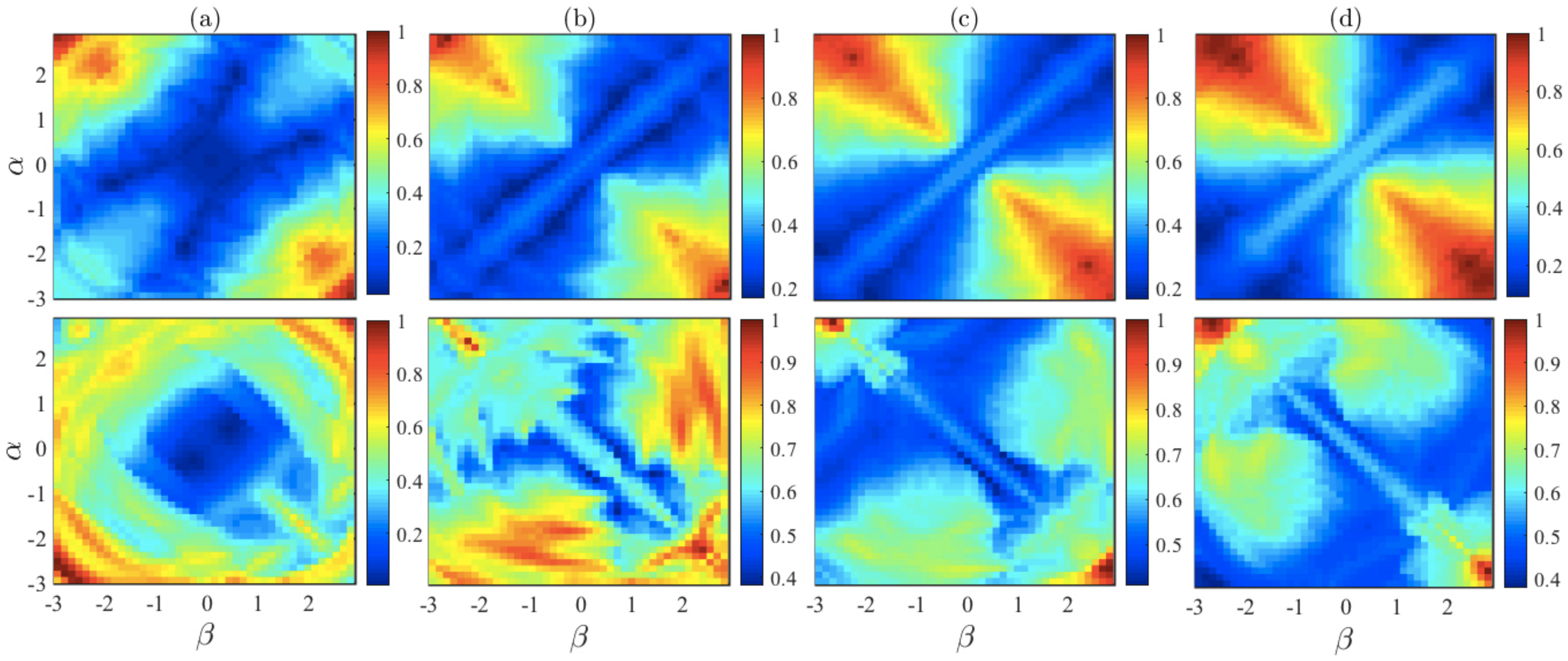}
\caption{\label{fig11} (Color online). Normalized maximum supercurrent in a ballistic Josephson junction vs the parameters of $\boldsymbol{\eta}_\text{so}^\text{b}$ RDSOI, $\alpha$ and $\beta$. The chemical potential increases from the left to the right column: (a) $\mu=\Delta$, (b) $\mu=5\Delta$, (c) $\mu=10\Delta$, and (d) $\mu=20\Delta$. In the top row, $h_x=0.75\Delta, h_y=0$ and bottom row $h_x=0, h_y=0.75\Delta$.}
\end{figure*}

The supercurrent-phase relation in the presence of $\boldsymbol{\eta}_\text{so}^\text{b}$ spin-orbit coupling, Eq. (\ref{so2}), when the junction is oriented along the $y$ direction:
\begin{subequations}\label{I_set_b1_appx}
\begin{eqnarray}
&& I(+\alpha,0, +h_x, 0)=I(0,+\beta, +h_x, 0)=I(-\alpha,0, -h_x, 0)=I(0,-\beta, -h_x, 0); \;\;\ \varphi_0^{\text{b},y}>0,\\
&& I(-\alpha,0, +h_x,0)=I(0,-\beta, +h_x,0)=I(+\alpha,0, -h_x,0)=I(0,+\beta, -h_x,0); \;\;\ \varphi_0^{\text{b},y}<0,
\end{eqnarray}
\end{subequations}
\begin{subequations}\label{I_set_b2_appx}
\begin{eqnarray}
&& I(+\alpha,+\beta, 0, +h_y)=I(-\alpha,-\beta, 0, +h_y)=I(+\alpha,+\beta, 0, -h_y)=I(-\alpha,-\beta, 0, -h_y); \;\;\ \varphi_0^{\text{b},y}=0,\\
&& I(+\alpha,-\beta, 0, +h_y)=I(-\alpha,+\beta, 0, +h_y)=I(+\alpha,-\beta, 0, -h_y)=I(-\alpha,+\beta, 0, -h_y); \;\;\ \varphi_0^{\text{b},y}=0,\\
&& I(+\alpha,-\beta, +h_x, 0)=I(-\alpha,+\beta, +h_x, 0)=I(+\alpha,-\beta, -h_x, 0)=I(-\alpha,+\beta, -h_x, 0); \;\;\ \varphi_0^{\text{b},y}=0,\\
&& I(+\alpha,+\beta, +h_x, 0)=I(-\alpha,-\beta, +h_x, 0)=I(+\alpha,+\beta, -h_x, 0)=I(-\alpha,-\beta, -h_x, 0); \;\;\ \varphi_0^{\text{b},y}=0,
\end{eqnarray}
\end{subequations}

\begin{subequations}\label{I_set_b3_appx}
\begin{eqnarray}
&& I(+\alpha,+\beta, +h_x, +h_y)=I(-\alpha,-\beta, -h_x, -h_y)=I(-\alpha,-\beta, -h_x, +h_y)=I(+\alpha,+\beta, +h_x, -h_y); \;\;\ \varphi_0^{\text{b},y}>0,\\
&& I(-\alpha,-\beta, +h_x, +h_y)=I(+\alpha,+\beta, -h_x, -h_y)=I(+\alpha,+\beta, -h_x, +h_y)=I(-\alpha,-\beta, +h_x, -h_y); \;\;\ \varphi_0^{\text{b},y}<0,
\end{eqnarray}
\end{subequations}

\begin{eqnarray}\label{I_set_b4_appx}
&& I(+\alpha,-\beta, +h_x, +h_y)=I(-\alpha,+\beta, +h_x, +h_y)=I(+\alpha,-\beta, -h_x, -h_y)=I(-\alpha,+\beta, -h_x, -h_y)=\nonumber\\ && I(+\alpha,-\beta, -h_x, +h_y)=I(-\alpha,+\beta, -h_x, +h_y)=I(+\alpha,-\beta, +h_x, -h_y)=I(-\alpha,+\beta, +h_x, -h_y); \;\;\  \varphi_0^{\text{b},y}=0,
\end{eqnarray}
\begin{subequations}\label{I_set_b5_appx}
\begin{eqnarray}
&& I(+\alpha,0,-h_x,+h_y)=I(+\alpha,0,-h_x,-h_y)=I(0,+\beta, -h_x, + h_y)=I(0,+\beta, -h_x, - h_y)=\nonumber\\ && I(0,-\beta, +h_x, + h_y)=I(0,-\beta, +h_x, - h_y)=I(-\alpha,0,+h_x,+h_y)=I(-\alpha,0,+h_x,-h_y); \;\;\ \varphi_0^{\text{b},y}>0,\\
&& I(+\alpha,0,+h_x,+h_y)=I(+\alpha,0,+h_x,-h_y)=I(0,+\beta, +h_x, +h_y)=I(0,+\beta, +h_x, - h_y)=\nonumber\\ && I(0,-\beta, -h_x, + h_y)=I(0,-\beta, -h_x, -h_y)=I(-\alpha,0,-h_x,+h_y)=I(-\alpha,0,-h_x,-h_y); \;\;\ \varphi_0^{\text{b},y}<0,
\end{eqnarray}
\end{subequations}

\section{Critical supercurrent in ballistic regime}\label{apnx_GF}
In this Appendix, we present the color-map profile of a critical supercurrent in a ballistic Josephson junction where a Zeeman field may introduce detrimental effect. In Figs. \ref{fig10} and \ref{fig11}, the SOI is $\boldsymbol{\eta}_\text{so}^\text{a}$ and $\boldsymbol{\eta}_\text{so}^\text{b}$, respectively. In the top row of Fig. \ref{fig10} and bottom row of Fig. \ref{fig11}, $h_x=0, h_y=0.75\Delta$ is set. Comparing with counterparts in Figs. \ref{fig6} and \ref{fig7}, it is apparent how an improper Zeeman field can remove the signature of persistent spin helices on critical supercurrents.

\twocolumngrid

\end{document}